\documentclass[aps,showpacs,tightenlines,nobalancelastpage,
floatfix,onecolumn]{revtex4}

\usepackage{amsmath}
\usepackage{amssymb}
\usepackage{graphicx}
\usepackage{dcolumn}
\usepackage{bm}

\graphicspath{{graphs-include/}}

\begin{document}

\title{Theory of Raman transitions in cavity QED}

\date{\today }

\author{A.~D.~Boozer}

\affiliation{
  Norman Bridge Laboratory of Physics 12-33,
  California Institute of Technology,
  Pasadena, CA 91125
}

\begin{abstract}
  We present two schemes for driving Raman transitions between the
  ground state hyperfine manifolds of a single atom trapped within a
  high-finesse optical cavity.
  In both schemes, the Raman coupling is generated by standing-wave
  fields inside the cavity, thus circumventing the optical access
  limitations that free-space Raman schemes must face in a cavity
  system.
  These cavity-based Raman schemes can be used to
  coherently manipulate both the internal and the motional degrees of
  freedom of the atom, and thus provide powerful tools for studying
  cavity quantum electrodynamics.
  We give a detailed theoretical analysis of each scheme, both for a
  three-level atom and for a multi-level cesium atom.
  In addition, we show how these Raman schemes can be used to cool the
  axial motion of the atom to the quantum ground state, and we perform
  computer simulations of the cooling process. 
\end{abstract}


\maketitle

\section{Introduction}

Systems consisting of a single atom coupled to a high-finesse optical
cavity are of fundamental importance to quantum optics and quantum
information science.
Such cavity QED systems have been experimentally implemented using
neutral atoms
\cite{ye99, mckeever03a, maunz04, sauer04, nussmann05a, miller05}
and ions
\cite{guth01, mundt02},
and have been the subject of numerous theoretical studies
\cite{englert91, haroche91, holland91, herkommer92, storey92,
  ren95, herkommer96, scully96}.
In particular, such systems play a key role in proposals for scalable
quantum computation
\cite{pellizzari95, duan04}
and distributed quantum networks
\cite{cirac97, briegel00}.
An important requirement for many of these proposals is the ability
to coherently control the internal and motional degrees of freedom of
the trapped atom, and Raman transitions provide the means for meeting
this requirement.

Raman transitions are powerful tools that have diverse applications in
atomic physics, including spectroscopy
\cite{dotesenko04},
precision measurement
\cite{clade06, gustavson97},
and coherent state manipulation
\cite{wineland03},
and have been used to coherently control the motional degrees of
freedom of trapped ions
\cite{monroe95, leibfried03}
and of neutral atoms trapped in optical lattices
\cite{hamann98, perrin98, vuletic98}.
Until recently, however, Raman transitions had not been incorporated
into cavity QED.
The practical challenge of implementing a free-space Raman scheme in
cavity QED lies in the presence of the cavity itself, which offers
limited optical access to the atom within, especially in the strong
coupling regime \cite{miller05}.
As first proposed in \cite{boozer05}, these optical access limitations
can be circumvented by implementing a cavity-based Raman scheme, in
which the Raman coupling is generated by standing-wave fields inside
the cavity.

In this paper, we provide the theoretical background behind two
such cavity-based schemes for driving Raman transitions between the
hyperfine ground state manifolds of a single atom optically trapped
within a high-finesse cavity.
Both schemes have been recently implemented and validated
experimentally,  and have been used to extend the trapping lifetime of
a near-resonantly driven atom
\cite{boca04},
to cool the axial motion of an atom to the quantum ground state of its
trapping potential
\cite{boozer06},
and to optically pump an atom into a specific Zeeman state
\cite{boozer07}.

The paper is organized as follows.
In section
\ref{sec:trapping},
we describe how a single atom is trapped inside the cavity by means of
an optical dipole trap.
In section
\ref{sec:raman-three-level},
we present the two schemes for driving Raman transitions in the
trapped atom.
We treat the atom using a three-level model, and show that the
internal and motional degrees of freedom of the atom can be described
using an effective Hamiltonian that has the same form for both
schemes.
In section
\ref{sec:quantize-axial},
we quantize the axial motion of the atom, and show how the Raman
couplings allow one to drive transitions that change the vibrational
quantum number for motion along the cavity axis.
We present both analytic results, which apply to atoms that are
sufficiently cold, and numerical results, which apply to atoms of
arbitrary temperature.
In section \ref{sec:raman-cesium}, we describe how the Raman couplings
are modified when we take into account
the multiplicity of levels in a physically realistic cesium atom.
We show that the Raman schemes drive transitions between individual
Zeeman states in the two hyperfine ground state manifolds, and we
derive the Rabi frequencies for these Zeeman transitions.
Finally, in section
\ref{sec:cooling},
we show how Raman transitions can be used to cool the axial motion
of the atom to the quantum ground state, and we present computer
simulations of the cooling process.

\section{Trapping an atom inside the cavity}
\label{sec:trapping}

The optical cavity we will be considering consists of two symmetric
mirrors separated by a distance $L$ (details of the cavity are given
in Appendix \ref{sec:cavity-mode-structure}).
An atom is trapped inside the cavity by means of a far off-resonance
trap (FORT), which is created by driving one of the
cavity modes with light that is red-detuned from a dipole transition
in the atom.
The red-detuned light forms a standing wave inside the cavity, and the
coupling of the atom to the light causes it to be attracted to the
points of maximum intensity in the standing wave.

The operation of the FORT can be understood by considering a simple
two-level model in which the atom has a single ground state $g$ and
a single excited state $e$.
Let $\omega_e$ denote the frequency of the $g-e$ transition, and let
$\omega_F$ denote the frequency of the FORT light.
We will assume that the FORT resonantly drives mode $n_F$ of the
cavity, so $\omega_F = 2\pi \nu_{FSR}\,n_F$, where $\nu_{FSR} = 1/2L$
is the free spectral range.
The Hamiltonian for the system is
\begin{eqnarray}
  H = \omega_e |e\rangle\langle e| +
  (\hat{\Omega} + \hat{\Omega}^\dagger)\,\cos \omega_F t,
\end{eqnarray}
where
\begin{eqnarray}
  \hat{\Omega} \equiv \Omega_F\,\psi_F(\vec{r})\,|g\rangle\langle e|.
\end{eqnarray}
Here $\vec{r}$ is the position of the atom, $\psi_F(\vec{r})$ is a
dimensionless quantity that characterizes the shape of the cavity
mode (see Appendix \ref{sec:cavity-mode-structure}), and $\Omega_F$ is
the Rabi frequency of the light at a point of maximum intensity.
We can simplify this Hamiltonian by making the rotating wave
approximation and then performing a unitary transformation to
eliminate the time dependence:
\begin{eqnarray}
  H = -\Delta_F |e\rangle\langle e| +
  \frac{1}{2}(\hat{\Omega} + \hat{\Omega}^\dagger),
\end{eqnarray}
where $\Delta_F \equiv \omega_L - \omega_e$ is the detuning of the FORT
from the atom
(note that because the FORT is red-detuned, $\Delta_F < 0$).
In the limit that the FORT is far-detuned
($|\Delta_F| \gg \Omega_F$), we can adiabatically eliminate the
excited state (see \cite{boozer05}) to obtain an effective Hamiltonian
for the ground state:
\begin{eqnarray}
  H_E = \frac{1}{4\Delta_F} \hat{\Omega}\hat{\Omega}^\dagger =
  U(\vec{r})\,|g\rangle\langle g|,
\end{eqnarray}
where
\begin{eqnarray}
  U(\vec{r}) \equiv -U_F\,|\psi_F(\vec{r})|^2
\end{eqnarray}
describes a trapping potential with depth
$U_F \equiv \Omega_F^2/4|\Delta_F|$.
For the experiments described in \cite{boca04,boozer06,boozer07}, the
power in the FORT beam is set such that
$U_F \simeq (2\pi)(50\,\mathrm{MHz})$.
To describe the shape of the potential, it is convenient to use a
cylindrical coordinate system centered on the cavity axis: we will let
$z$ and $\rho$ denote the axial and radial coordinates of the atom,
where the cavity mirrors are located at $z=0$ and $z=L$.
Using equation (\ref{eqn:mode-shape-general}) to substitute for
the mode shape $\psi_F(\vec{r})$ in this coordinate system, we find
that
\begin{eqnarray}
  U(\vec{r}) = -U_F\,e^{-2\rho^2/w_F^2}\,\sin^2 k_F z,
\end{eqnarray}
where $k_F \equiv \pi n_F/L$ is the wavenumber for the FORT mode
$n_F$.
The minima of the potential occur at the points $\rho=0$, $z = z_r$,
where $z_r$ is defined such that
\begin{eqnarray}
  \label{eqn:fort-well-locations}
  k_F z_r = \pi(r + 1/2).
\end{eqnarray}
Since $0 < z_r < L$, we find that $r = 1, ..., n_F-1$; thus, there are
$n_F$ distinct FORT wells in which an atom can be trapped.
Let us assume that the atom is trapped in FORT well $r$.
It is convenient to define a coordinate $x = z - z_r$ that gives the
axial displacement of the atom from the potential minimum of this
well.
We can then express the trapping potential as
\begin{eqnarray}
  U(\vec{r}) = -U_F\,e^{-2\rho^2/w_F^2}\,\cos^2 k_F x.
\end{eqnarray}
Near the bottom of the well the potential can be approximated as
harmonic,
\begin{eqnarray}
  U(\vec{r}) \simeq
  -U_F + \frac{1}{2}m\omega_r^2 \rho^2 +
  \frac{1}{2}m\omega_a^2 x^2,
\end{eqnarray}
where $\omega_r$ and $\omega_a$, the radial and axial vibrational
frequencies, are given by
\begin{eqnarray}
  \frac{1}{2}m\omega_r^2 = 2 U_F/w_F^2,
  \qquad
  \frac{1}{2}m\omega_a^2 = U_F k_F^2.
\end{eqnarray}
The corresponding periods $2\pi/\omega_r$ and $2\pi/\omega_a$
characterize the timescales for radial and axial motion.
For the experiments described in \cite{boca04,boozer06,boozer07},
the vibrational frequencies are
$\omega_r \simeq (2\pi)(5\,\mathrm{kHz})$
and $\omega_a \simeq (2\pi)(500\,\mathrm{kHz})$, so the timescale for
radial motion is much longer than the timescale for axial motion.
We will be interested in describing the evolution of system over
timescales that are short compared to the timescale for radial motion,
but not necessarily short compared to the timescale for axial motion.
For such timescales we can view the atom as being radially stationary,
and take $\rho$ to be a constant parameter that enters into the
potential for the axial motion.
Thus, dropping a constant term, we can express the potential for the
axial motion as
\begin{eqnarray}
  U(\vec{r}) = U_\rho \sin^2 k_F x,
\end{eqnarray}
where $U_\rho \equiv U_F\,e^{-2\rho^2/w_F^2}$ is the axial trap depth
at radial coordinate $\rho$.

\section{Raman coupling for a three-level model}
\label{sec:raman-three-level}

\subsection{Effective Hamiltonian}
\label{ssec:raman-three-level-effective-hamiltonian}

To show how a Raman coupling can be generated in a trapped
atom, let us now consider a three-level model in which the atom has
two ground states $a$ and $b$, which correspond to the ground state
hyperfine manifolds of a multi-level atom, and a single excited state
$e$.
The excited state has energy $\omega_e$, and the ground states $a$ and
$b$ have energies $-\Delta_{HF}/2$ and $+\Delta_{HF}/2$, where
$\Delta_{HF}$ is the ground state hyperfine splitting.
The Hamiltonian for the atom is
\begin{eqnarray}
  \label{eqn:h0-three-level}
  H_0 = \omega_e |e\rangle\langle e| + \frac{1}{2}\Delta_{HF}\sigma_z,
\end{eqnarray}
where
\begin{eqnarray}
  \sigma_z \equiv |b\rangle\langle b| - |a\rangle\langle a|.
\end{eqnarray}

\begin{figure}
  \centering
  \includegraphics[scale=0.8]{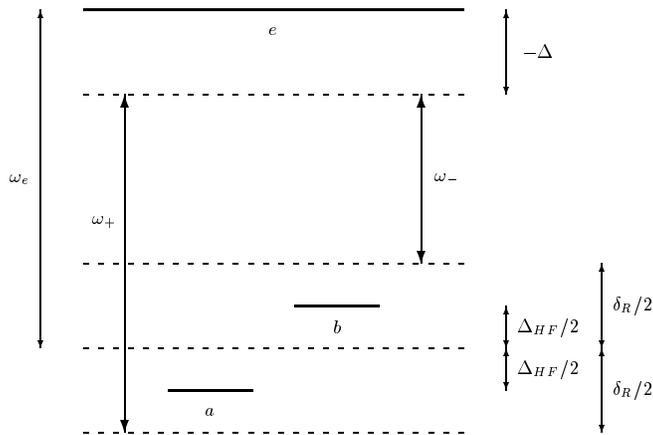}
  \caption{
    Level diagram for the three-level atom.
    Shown are the ground states $a$ and $b$, the excited state $e$,
    and the pair of optical fields at frequencies $\omega_+$ and
    $\omega_-$.
  }
  \label{fig:raman-config-levels}
\end{figure}

We can generate a Raman coupling between the two ground states by
driving one of the cavity modes with a pair of beams that are tuned
into Raman resonance with the ground state hyperfine splitting of the
atom.
Let us denote the optical frequencies of these beams by
$\omega_\pm = \omega_L \pm \delta_R/2$,
where $\omega_L$ is the average frequency and $\delta_R$ is the
frequency difference (see Figure \ref{fig:raman-config-levels}).
Also, let us define a parameter $\delta = \delta_R - \Delta_{HF}$
that gives the Raman detuning of the beams; we will assume that the
beams are tuned close to Raman resonance, so $|\delta| \ll
\Delta_{HF}$.
The beams generate standing-wave fields inside the cavity, and the
coupling of the atom to these fields is described by the Hamiltonian
\begin{eqnarray}
  \label{eqn:raman-pair-three-level}
  H_R =
  (\hat{\Omega}_+ + \hat{\Omega}_+^\dagger) \cos\omega_+ t +
  (\hat{\Omega}_- + \hat{\Omega}_-^\dagger) \cos\omega_- t,
\end{eqnarray}
where
\begin{eqnarray}
  \hat{\Omega}\pm \equiv \Omega_\pm\,\psi(\vec{r})\,A,
\end{eqnarray}
and
\begin{eqnarray}
  \label{eqn:a-three-level}
  A \equiv (|a\rangle + |b\rangle) \langle e|
\end{eqnarray}
is an atomic lowering operator.
Here $\vec{r}$ is the position of the atom, $\psi(\vec{r})$ is a
dimensionless quantity that characterizes the shape of the driven
mode (see Appendix \ref{sec:cavity-mode-structure}), and $\Omega_\pm$
are the Rabi frequencies of the fields at a point of maximum
intensity.
For simplicity, we have assumed that the $a-e$ and $b-e$ transitions
couple to the light fields with equal strength.

The total Hamiltonian for the system is $H = H_0 + H_R$.
We can simplify this Hamiltonian by making the rotating wave
approximation and then performing a unitary transformation:
\begin{eqnarray}
  H =
  -\Delta |e\rangle\langle e| + \frac{1}{2}\Delta_{HF} \sigma_z +
  \hat{B} + \hat{B}^\dagger,
\end{eqnarray}
where
\begin{eqnarray}
  \hat{B} \equiv
  \frac{1}{2}(\hat{\Omega}_+\,e^{i\delta_R t/2} +
  \hat{\Omega}_-\,e^{-i\delta_R t/2}),
\end{eqnarray}
and $\Delta \equiv \omega_L - \omega_e$ describes the overall detuning
of the optical fields from the excited state.
We will assume that the fields are far-detuned from the atom
($|\Delta| \gg \Omega_\pm$), so we can further simplify the Hamiltonian
by adiabatically eliminating the excited state to obtain an effective
Hamiltonian for the ground states:
\begin{eqnarray}
  H_E
  & = &
  \frac{1}{2}\Delta_{HF} \sigma_z +
  \frac{1}{\Delta} \hat{B} \hat{B}^\dagger, \\
  & = &
  \frac{1}{2}\Delta_{HF} \sigma_z -
  V_E\,|\psi(\vec{r})|^2 - V_E\,|\psi(\vec{r})|^2\,\sigma_x +
  \Omega_E\,|\psi(\vec{r})|^2\cos\delta_R t +
  \Omega_E\,|\psi(\vec{r})|^2\,\sigma_x \cos\delta_R t,
\end{eqnarray}
where
\begin{eqnarray}
  V_E \equiv (\Omega_+^2 + \Omega_-^2)/4|\Delta|,
  \qquad
  \Omega_E \equiv \Omega_+\,\Omega_-/2\Delta,
\end{eqnarray}
and
\begin{eqnarray}
  \sigma_x \equiv |a\rangle \langle b| + |b\rangle \langle a|.
\end{eqnarray}
We will assume that the optical fields are weak enough that
$V_E, \Omega_E \ll \Delta_{HF}$, so the first term of $H_E$ dominates
and to a good approximation the eigenstates of $H_E$ are $|a\rangle$
and $|b\rangle$.
The second term of $H_E$ describes a state-independent level shift,
which is analogous to the FORT potential we derived in the previous
section.
The third term of $H_E$ gives a state-dependent correction to the
level shift, which is of order $V_E^2/\Delta_{HF}$ and may therefore
be neglected.
The fourth term of $H_E$ describes a modulation of the
state-independent level shift at frequency $\delta_R$.
Because we have assumed that the system is tuned near to
Raman resonance, $\delta_R$ is of the same order as the hyperfine
splitting $\Delta_{HF}$.
For cesium, the atom used in the experiments described in
\cite{boca04,boozer06,boozer07}, the hyperfine splitting is
$\Delta_{HF} = (2\pi)(9.2\,\mathrm{GHz})$, which is much larger than
the harmonic frequencies $\omega_r$ and $\omega_a$ that characterize
the timescales for atomic motion.
Thus, over the motional timescales the fourth term of $H_E$ averages
to zero and may also be neglected.
After making these approximations, we are left with
\begin{eqnarray}
  H_E = \frac{1}{2}\Delta_{HF} \sigma_z -
  V_E\,|\psi(\vec{r})|^2 +
  \Omega_E\,|\psi(\vec{r})|^2 \sigma_x\cos\delta_R t.
\end{eqnarray}
We can further simplify the effective Hamiltonian by
making the rotating wave approximation and then performing a unitary
transformation to eliminate the time-dependence:
\begin{eqnarray}
  \label{eqn:effective-hamiltonian}
  H_E = -\frac{\delta}{2}\sigma_z  - V_E\,|\psi(\vec{r})|^2 +
  \frac{1}{2}\,\Omega_E\,|\psi(\vec{r})|^2\,\sigma_x.
\end{eqnarray}
This Hamiltonian describes an effective two-level atom with ground
state $a$ and excited state $b$, which is driven by a classical field
with Rabi frequency $\Omega_E\,|\psi(\vec{r})|^2$ and detuning
$\delta$.
We will now use this effective Hamiltonian to describe two schemes for
driving Raman transitions in a trapped atom.

\subsection{FORT-Raman configuration}
\label{sec:fort-raman-three-level}

In the first scheme, which we will call the FORT-Raman
configuration, the FORT itself forms one leg of a Raman pair.
To form the other leg of the pair we add a much weaker beam, which we
will call the Raman beam, that
drives the same cavity mode as the FORT but is detuned from the cavity
resonance by $\delta_R$.
This configuration was first proposed in \cite{boozer05}, and formed
the basis of the optical pumping scheme described in \cite{boozer07}
and of the cooling scheme used in \cite{boca04} to extend the lifetime
of a trapped atom.
Let us denote the optical frequencies of the FORT and Raman beams
by $\omega_F$ and $\omega_R = \omega_F + \delta_R$, and their maximum
Rabi frequencies inside the cavity by $\Omega_F$ and $\Omega_R$.
We can then apply the results of section
\ref{ssec:raman-three-level-effective-hamiltonian}
to obtain an effective Hamiltonian $H_E$ that describes the
FORT-Raman pair.
We will assume that the Raman beam is blue-detuned from the FORT
($\delta_R > 0$), so we take the $\Omega_+$ field to be the Raman
field and the $\Omega_-$ field to be the FORT field; if the Raman beam
is red-detuned from the FORT ($\delta_R < 0$) then we reverse these
identifications.

As was discussed in section \ref{sec:trapping}, we treat the atom as
being radially stationary, and consider only the axial motion.
Thus, the total Hamiltonian for the atom is given by
\begin{eqnarray}
  H = \frac{p^2}{2m} + H_E.
\end{eqnarray}
Here $p$ is the momentum of the atom in the axial direction,
and $H_E$ is given by equation (\ref{eqn:effective-hamiltonian}) with
$\Omega_E = \Omega_F\Omega_R/2\Delta_F$ and $V_E = U_F + U_R$, 
where
$U_F = \Omega_F^2/4|\Delta_F|$ and $U_R = \Omega_R^2/4|\Delta_F|$ are
the maximum level shifts due to the FORT and Raman fields
individually, and $\Delta_F = \omega_F - \omega_e$ is the overall
detuning of the FORT and Raman beams from the atom.
It is convenient to express $H$ as $H = H_{ext} + H_{int}$, where
\begin{eqnarray}
  \label{eqn:h-ext-fr}
  H_{ext} = \frac{p^2}{2m} - V_E\,|\psi_F(\vec{r})|^2
\end{eqnarray}
describes the axial motion of the atom, and
\begin{eqnarray}
  H_{int} = 
  -\frac{\delta}{2} \sigma_z +
  \frac{1}{2}\,\Omega_E\,|\psi_F(\vec{r})|^2\,\sigma_x
\end{eqnarray}
describes the internal state of the atom.
Typically the powers of the FORT and Raman beams are such
that $\Omega_E \simeq (2\pi)(200\,\mathrm{kHz})$ and
$U_F \simeq (2\pi)(50\,\mathrm{MHz})$, and for these values
$U_R/U_F \sim (\Omega_E/U_F)^2 \simeq 2 \times 10^{-5}$.
Thus, we can neglect the level shift of the Raman beams and
approximate $H_{ext}$ as
\begin{eqnarray}
  H_{ext} = \frac{p^2}{2m} + U(\vec{r}),
\end{eqnarray}
where $U(\vec{r}) = -U_F\,|\psi_F(\vec{r})|^2$ is the FORT trapping
potential.
As in section \ref{sec:trapping}, we will assume that the atom is
trapped in FORT well $r$, and define a coordinate $x = z - z_r$ that
gives the axial displacement of the atom from the well minimum.
Note that
\begin{eqnarray}
  |\psi_F(\vec{r})|^2 =
  e^{-2\rho^2/w_F^2}\,\sin^2 k_F z =
  e^{-2\rho^2/w_F^2}\,\cos^2 k_F x,
\end{eqnarray}
where we have substituted for the FORT mode shape
$\psi_F(\vec{r})$ using equation (\ref{eqn:mode-shape-general})
and for $k_F z_r$ using equation (\ref{eqn:fort-well-locations}).
Thus, we can express the internal Hamiltonian as
\begin{eqnarray}
  H_{int} =
  -\frac{\delta}{2} \sigma_z  +
  \frac{1}{2}\,\Omega_\rho \cos^2 k_F x\,\sigma_x,
\end{eqnarray}
where $\Omega_\rho \equiv \Omega_E\,e^{-2\rho^2/w_F^2}$ is the
effective Rabi frequency for an atom at radial coordinate $\rho$.

\subsection{Raman-Raman configuration}
\label{sec:raman-raman-three-level}

For the second configuration, which we will call the
Raman-Raman configuration, the FORT drives mode $n_F$ of the cavity
at frequency $\omega_F$, and a pair of Raman beams drives mode $n_R$ of
the cavity at frequency $\omega_R$.
This configuration was used to perform the ground-state cooling
described in \cite{boozer06}.
We will assume that the Raman beams have equal powers and are tuned
symmetrically about the cavity resonance, so the frequencies of these
beams can be expressed as $\omega_R \pm \delta_R/2$.
Let $\Omega_F$ denote the maximum Rabi frequency for the FORT beam,
and let $\Omega_R$ denote the maximum Rabi frequency for one of the
Raman beams.
Using the results of section
\ref{ssec:raman-three-level-effective-hamiltonian}, we can
describe the coupling of the atom to the pair of Raman fields in terms
of an effective Hamiltonian $H_E$.

As was discussed in section \ref{sec:trapping}, we treat the atom as
being radially stationary, and consider only the axial motion.
Thus, the total Hamiltonian for the system is
\begin{eqnarray}
  H = \frac{p^2}{2m} + U(\vec{r}) + H_E.
\end{eqnarray}
The first term describes the kinetic energy of the atom due to axial
motion, and the second term describes the FORT potential
$U(\vec{r}) = -U_F\,|\psi_F(\vec{r})|^2$, where
$U_F = \Omega_F^2/4|\Delta_F|$ is the FORT depth.
The third term is given by equation (\ref{eqn:effective-hamiltonian})
with $V_E = 2 U_R$,  $U_R = \Omega_R^2/4|\Delta_R|$, and
$\Omega_E = \Omega_R^2/2\Delta_R$, where
$\Delta_R \equiv \omega_R - \omega_e$ is the overall detuning of the
Raman pair from the atom.
It is convenient to express $H$ as $H = H_{ext} + H_{int}$, where
\begin{eqnarray}
  \label{eqn:h-ext-raman-raman}
  H_{ext} = \frac{p^2}{2m} + U(\vec{r}) - V_E|\psi_R(\vec{r})|^2
\end{eqnarray}
describes the motion of the atom, and
\begin{eqnarray}
  H_{int} = 
  -\frac{\delta}{2} \sigma_z +
  \frac{1}{2}\,\Omega_E\,|\psi_R(\vec{r})|^2\,\sigma_x
\end{eqnarray}
describes the internal state of the atom.

Note that because the FORT and Raman beams drive different cavity
modes, the registration of the standing waves corresponding to the two
beams depends on the axial position of the atom, and therefore on the
particular FORT well in which the atom is trapped.
This is in contrast to the FORT-Raman configuration, for which
the FORT and Raman standing waves are always perfectly registered.
One consequence of the well-dependence of the registration is that the
level shift $V_E$ due to the Raman pair distorts different FORT wells
in different ways.
We calculate this effect in Appendix \ref{sec:potential-distortion},
but for now we note that for the typical parameters
$\Omega_E \simeq (2\pi)(200\,\mathrm{kHz})$,
$U_F \simeq (2\pi)(50\,\mathrm{MHz})$
the ratio of the level shifts due
to the FORT and Raman beams is
$U_R/U_F = \Omega_E/U_F \simeq 4 \times 10^{-3}$.
Thus $V_E \ll U_F$, so we can neglect the $V_E$ term and approximate
$H_{ext}$ as
\begin{eqnarray}
  H_{ext} = \frac{p^2}{2m} + U(\vec{r}).
\end{eqnarray}
As in section \ref{sec:trapping},
we will assume that the atom is trapped in FORT well $r$, and define a
coordinate $x = z - z_r$ that gives the axial displacement of the atom
from the well minimum.
Note that
\begin{eqnarray}
  \label{eqn:raman-intensity-profile}
  |\psi_R(\vec{r})|^2 =
  e^{-2\rho^2/w_R^2}\,\sin^2 k_R z =
  e^{-2\rho^2/w_R^2}\,\cos^2(k_R x + \alpha),
\end{eqnarray}
where $\alpha = (k_R - k_F) z_r$ is the phase difference between the
FORT and Raman beams at the bottom of FORT well $r$.
Here we have substituted for the FORT mode shape
$\psi_F(\vec{r})$ using equation (\ref{eqn:mode-shape-general})
and for $k_F z_r$ using equation (\ref{eqn:fort-well-locations}).
We will assume that the FORT and Raman beams drive nearby modes of the
cavity, so $|n_R - n_F| \ll n_R, n_F$.
In this limit $(k_R - k_F) x \ll 1$ and $w_R \simeq w_F$, so we can
approximate equation (\ref{eqn:raman-intensity-profile}) as
\begin{eqnarray}
  |\psi_R(\vec{r})|^2 =
  e^{-2\rho^2/w_F^2}\,\cos^2 (k_F x + \alpha).
\end{eqnarray}
Thus, we can express the internal Hamiltonian as
\begin{eqnarray}
  H_{int} =
  -\frac{\delta}{2} \sigma_z  +
  \frac{1}{2}\,\Omega_\rho\cos^2 (k_F x + \alpha)\,\sigma_x,
\end{eqnarray}
where $\Omega_\rho \equiv \Omega_E\,e^{-2\rho^2/w_F^2}$ is the
effective Rabi frequency for an atom at radial coordinate $\rho$.

\subsection{Summary of Raman schemes}

We have shown that that for both the FORT-Raman and the Raman-Raman
configurations the Hamiltonian has the form
$H = H_{ext} + H_{int}$, where
\begin{eqnarray}
  \label{eqn:h-ext}
  H_{ext} & = &
  \frac{p^2}{2m} + U_\rho\,\sin^2 k_F x, \\
  \label{eqn:h-int}
  H_{int} & = &
  -\frac{\delta}{2} \sigma_z  +
  \frac{1}{2}\,\Omega_\rho\,\cos^2 (k_F x + \alpha)\,\sigma_x.
\end{eqnarray}
For the FORT-Raman configuration $\alpha = 0$, and for the
Raman-Raman configuration $\alpha = (k_R - k_F) z_r$ for an atom
trapped in FORT well $r$.
The Hamiltonian $H_{ext}$ describes the motion of the atom in the
axial potential and is independent of the atomic state, and the
Hamiltonian $H_{int}$ describes a Raman coupling between the ground
states that depends on the axial position of the atom.

For either the FORT-Raman or the Raman-Raman configuration, the Raman
coupling described by $H_{int}$ can be turned off by turning off one
of the beams in the Raman pair (note that for FORT-Raman configuration
the FORT beam must always be on in order to maintain the trapping
potential, so for this configuration one must turn off the Raman beam).
Alternatively, one can keep the beams in the Raman pair on at all
times and turn off the Raman coupling by tuning the pair out of
Raman resonance ($|\delta| \gg \Omega_\rho$).
As we have seen, there is a small level shift due to the Raman beams,
which we neglected when writing down the above expression for
$H_{ext}$, and this detuning-based method has the advantage that these
level shifts are always present regardless of whether the Raman
coupling is on or off.
With the first method, these level shifts cause a slight change in the
trapping potential whenever the Raman coupling is turned on or off,
which could potentially heat the atom or cause other problems.

\section{Quantization of axial motion}
\label{sec:quantize-axial}

For many applications, such as Raman sideband cooling, it is necessary
to quantize the axial motion; that is, to treat the axial position $x$
and momentum $p$ as quantum operators.
We first show how this is achieved for cold atoms, and then discuss
some numerical results that apply to atoms of arbitrary temperature.

\subsection{Approximate form of the Hamiltonian for cold atoms}

For cold atoms the axial trapping potential is nearly harmonic, where
the harmonic frequency $\omega$ is given by
\begin{eqnarray}
  \omega = (2 U_\rho/m)^{1/2}\,k_F = \omega_a\,e^{-\rho^2/w_F^2}.
\end{eqnarray}
We can quantize the axial motion by introducing phonon creation and
annihilation operators operators $b^\dagger$ and $b$, which are
related to $x$ and $p$ by
\begin{eqnarray}
  x = (2m\omega)^{-1/2}\,(b + b^\dagger),
  \qquad
  p = -i(m\omega/2)^{1/2}\,(b - b^\dagger).
\end{eqnarray}
From these relations, we find that
\begin{eqnarray}
  \label{eqn:kf-x}
  k_F x = \eta(b + b^\dagger),
\end{eqnarray}
where $\eta$, the Lamb-Dicke parameter, is given by
$\eta \equiv (2m\omega)^{-1/2} \omega_F$.
Note that because $\omega$ depends on the radial coordinate $\rho$,
the Lamb-Dicke parameter also depends on $\rho$.

If the atoms are sufficiently cold, we can obtain a reasonable
approximation to $H = H_{int} + H_{ext}$ by expanding in $\eta$ and
retaining terms only up to second order; from equations
(\ref{eqn:h-ext}), (\ref{eqn:h-int}), and (\ref{eqn:kf-x}), we find
that
\begin{eqnarray}
  H_{ext} & = &
  \omega (1/2 + b^\dagger b) - \omega(\eta^2/12)(b + b^\dagger)^4, \\
  H_{int} & = &
  -\frac{\delta}{2} \sigma_z  +
  \frac{1}{2}\Omega_\rho\,((1/2)(1 + \cos 2\alpha) -
  \eta (b + b^\dagger)\sin 2\alpha -
  \eta^2 (b + b^\dagger)^2 \cos 2\alpha)\,\sigma_x.
\end{eqnarray}
It is convenient to form a basis of states
$\{|a,n\rangle, |b,n\rangle\}$ for the system by
taking tensor products of the internal states $|a\rangle$ and
$|b\rangle$ with the motional Fock states $\{ |n\rangle \}$.
To order $\eta^2$ these product states are eigenstates of $H_{ext}$,
where pairs of states $|a,n\rangle$, $|b,n\rangle$ with the same
vibrational quantum number $n$ are degenerate and have energy
\begin{eqnarray}
  E_n =
  \langle a, n |H_{ext}|a, n\rangle =
  \langle b, n |H_{ext}|b, n\rangle =
  \omega(1/2 + n) - \omega(\eta^2/4)(1 + 2n + 2n^2).
\end{eqnarray}
The Raman coupling described by $H_{int}$ drives transitions between
the product states.
By taking matrix elements of the Raman coupling, we find that state
$|a,n\rangle$ is coupled to states $|b,n\rangle$, $|b,n\pm 1\rangle$,
and $|b, n \pm 2\rangle$, where the Rabi frequencies for these
transitions are given by
\begin{eqnarray}
  \Omega_{n \rightarrow n}/\Omega_\rho
  & = &
  1/2 + (1/2 - \eta^2(2n+1))\,\cos 2\alpha, \\
  \Omega_{n \rightarrow n \pm 1}/\Omega_\rho
  & = &
  -\eta\sqrt{n \pm 1}\,\sin 2\alpha, \\
  \Omega_{n \rightarrow n \pm 2}/\Omega_\rho
  & = &
  -\eta^2\sqrt{n \pm 1}\sqrt{n \pm 2}\,\cos 2\alpha.
\end{eqnarray}
Note that $\Delta n = \pm 1$ transitions are suppressed relative
to $\Delta n = 0$ transitions by $\sim \eta\sqrt{n}$, and
$\Delta n = \pm 2$ transitions are suppressed relative to
$\Delta n = 0$ transitions by $\sim \eta^2\,n$.
To resonantly drive the $n \rightarrow n$ transition we set
$\delta = 0$, and to resonantly drive the $n \rightarrow n \pm 1$ and
$n \rightarrow n \pm 2$ transitions we set
$\delta = \delta_{n \rightarrow n\pm 1}$ and
$\delta = \delta_{n \rightarrow n \pm 2}$, where
\begin{eqnarray}
  \delta_{n\rightarrow n \pm 1} =
  E_{n\pm 1} - E_n \simeq
  \pm \delta_n,
  \qquad
  \delta_{n\rightarrow n \pm 2} =
  E_{n\pm 2} - E_n \simeq
  \pm 2\delta_n,
\end{eqnarray}
and $\delta_n \equiv \omega - \eta^2 \omega\,n$.
For a harmonic trap $\delta_n = \omega$ and the frequencies of these
transitions are independent of $n$, but because the FORT is shallower
than its harmonic approximation, $\delta_n$ decreases with increasing
$n$.

Recall that for the FORT-Raman configuration $\alpha=0$, whereas for
the Raman-Raman configuration the value of $\alpha$ depends on the
FORT well in which the atom is trapped.
Thus, in the FORT-Raman configuration the Rabi frequencies are the
same for all the FORT wells, whereas in the Raman-Raman configuration
the Rabi frequencies vary from well to well.
Also, note that in the FORT-Raman configuration the $\Delta n = \pm 1$
transitions are always forbidden.
This follows from symmetry considerations: since the trapping
potential is symmetric under $x \rightarrow -x$ the motional
eigenstates are also parity eigenstates, and since the Raman coupling
is symmetric under $x \rightarrow -x$ it cannot couple an even parity
state to an odd parity state.

\subsection{Numerical results}

\begin{figure}[h]
  \centering
  \includegraphics[scale=0.6]{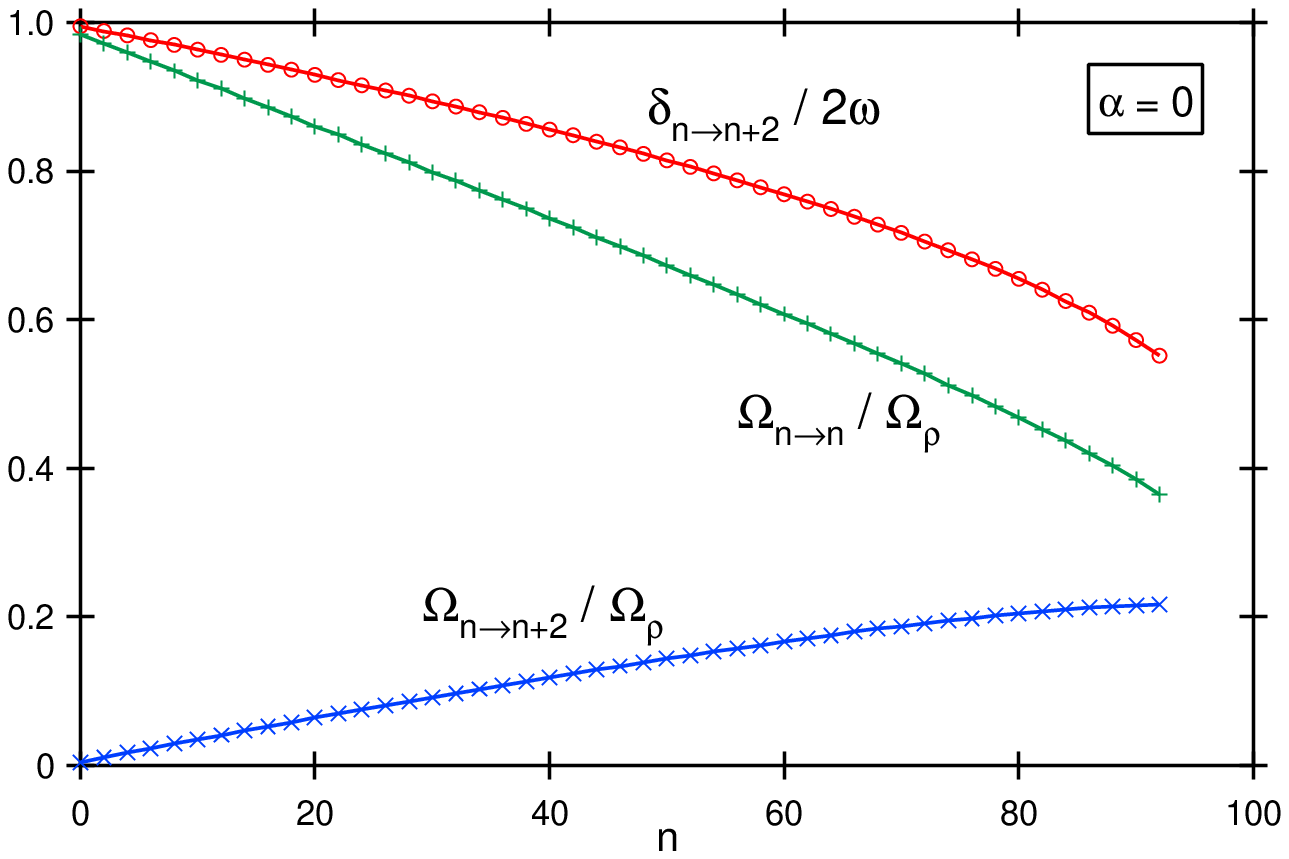}
  \includegraphics[scale=0.6]{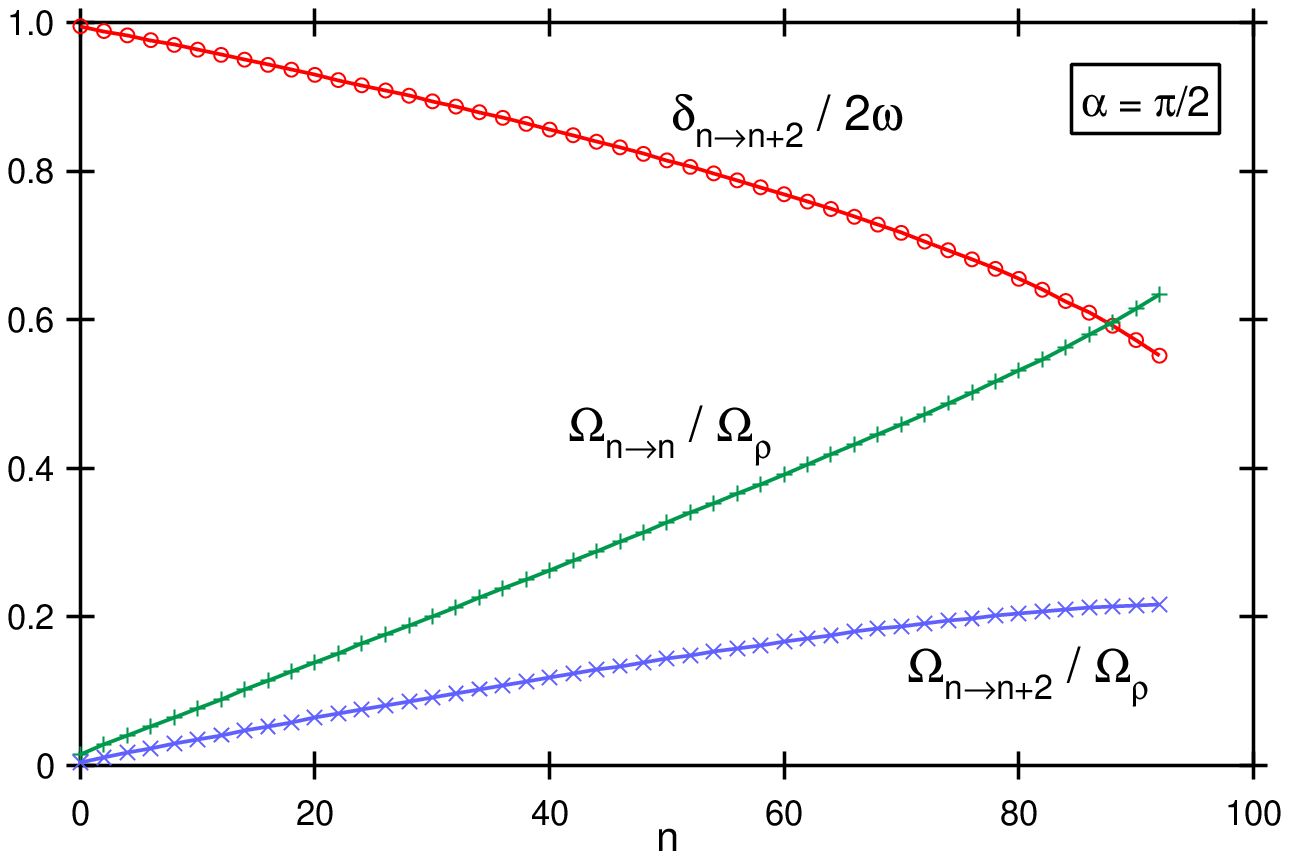}
  \caption{
    Rabi frequencies and detunings for FORT wells with
    $\alpha = 0, \pi/2$:
    green curve is $\Omega_{n\rightarrow n}/\Omega_\rho$,
    blue curve is $\Omega_{n\rightarrow n+2}/\Omega_\rho$,
    red curve is $\delta_{n\rightarrow n+2}/2\omega$.
  }
  \label{fig:motional-parameters-1}
\end{figure}

\begin{figure}
  \centering
  \includegraphics[scale=0.6]{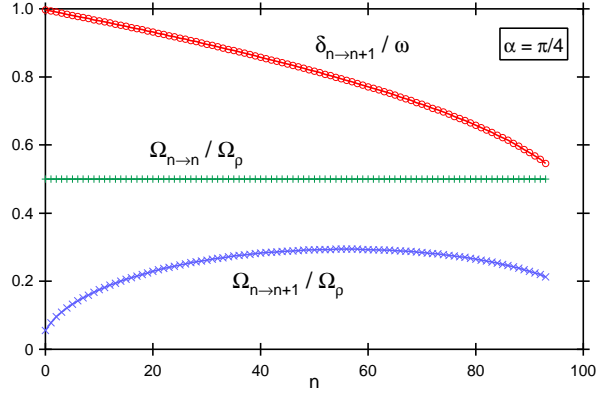}
  \caption{
    Rabi frequencies and detunings for FORT wells with
    $\alpha = \pi/4$:
    green curve is $\Omega_{n\rightarrow n}/\Omega_\rho$,
    blue curve is $\Omega_{n\rightarrow n+1}/\Omega_\rho$,
    red curve is $\delta_{n\rightarrow n+1}/\omega$.
  }
  \label{fig:motional-parameters-2}
\end{figure}

The harmonic approximation we described in the previous section only
applies to Fock states $|n\rangle$ for which $\eta \sqrt{n} \ll 1$.
For higher-lying energy levels, we can calculate the Rabi frequencies
and detunings for the various motional transitions by numerically
solving the time-independent Schr{\"o}dinger equation for $H_{ext}$.
This provides us with a set of motional eigenstates
$\{ |\psi_n \rangle \}$ and eigenvalues $\{ \bar{E}_n \}$.
Using the motional eigenstates, we can take matrix elements of the
Raman coupling described by $H_{int}$ to calculate the
Rabi frequencies for different motional transitions:
\begin{eqnarray}
  \Omega_{n \rightarrow n}/\Omega_\rho
  & = &
  \langle \psi_n|\cos^2(k_F x + \alpha) | \psi_n\rangle, \\
  \Omega_{n \rightarrow n \pm 1}/\Omega_\rho
  & = &
  \langle \psi_{n\pm 1}|\cos^2(k_F x + \alpha) | \psi_n\rangle, \\
  \Omega_{n \rightarrow n \pm 2}/\Omega_\rho
  & = &
  \langle \psi_{n \pm 2}|\cos^2(k_F x + \alpha) | \psi_n\rangle.
\end{eqnarray}
Note that $H_{int}$ can also drive higher-order $n$-changing transitions,
but the matrix elements for these transitions are quite small and we
will not consider them here.
From the energy eigenvalues, we can determine the detunings for the
$n \rightarrow n \pm 1$ and $n \rightarrow n \pm 2$ transitions:
\begin{eqnarray}
  \delta_{n\rightarrow n \pm 1} =
  \bar{E}_{n\pm 1} - \bar{E}_n,
  \qquad
  \delta_{n\rightarrow n \pm 2} =
  \bar{E}_{n\pm 2} - \bar{E}_n.
\end{eqnarray}
The numerically-determined Rabi frequencies and detunings are shown
in
Figure \ref{fig:motional-parameters-1} for $\alpha = 0$ and
$\alpha = \pi/2$, and
in Figure \ref{fig:motional-parameters-2} for $\alpha = \pi/4$.
Note that for $\alpha = 0$ and $\alpha = \pi/2$ the $\Delta n = \pm 1$
transitions are forbidden, so we only plot the Rabi frequencies and
detunings for the $\Delta n = 0$ and $\Delta n = \pm 2$ transitions,
and for $\alpha = \pi/4$ the $\Delta n = \pm 2$ transitions are
forbidden, so we only plot the Rabi frequencies and detunings for the
$\Delta n = 0$ and $\Delta n = \pm 1$ transitions.
For these graphs the Lamb-Dicke parameter is taken to be
$\eta = 0.05$, which is the value relevant for the experiments
described in \cite{boca04,boozer06,boozer07}.

\section{Raman coupling for cesium}
\label{sec:raman-cesium}

\subsection{Effective Hamiltonian}

\begin{figure}
  \centering
  \includegraphics[scale=1.0]{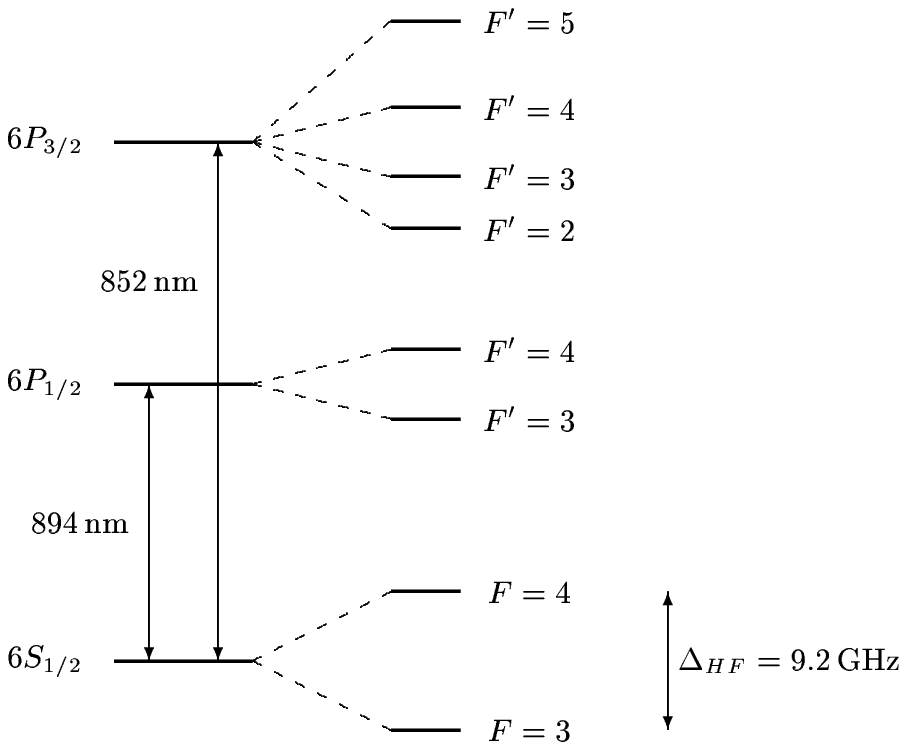}
  \caption{
    \label{fig:cesium-spectrum}
    Level diagram for cesium.}
\end{figure}

So far we have discussed the FORT-Raman and Raman-Raman schemes in the
context of a simple three-level model.
In this section, we show how these schemes are modified when we take
into account the multiplicity of levels in a physically realistic
alkali atom, using cesium as an example.

A level diagram for cesium is shown in Figure
\ref{fig:cesium-spectrum}; the levels relevant to our considerations
include the ground state hyperfine manifolds $6S_{1/2}, F=3$ and
$6S_{1/2}, F=4$, which correspond to ground states $a$ and $b$ of the
three-level model, and the excited state manifolds
$6P_{3/2}$ and $6P_{1/2}$, which correspond to the excited state $e$
of the three-level model.
The Hamiltonian for a free cesium atom is
\begin{eqnarray}
  H_0 = \sum_e \omega_e |e\rangle\langle e| +
  \frac{1}{2}\Delta_{HF}(P_4 - P_3),
\end{eqnarray}
where the sum is taken over all the states $e$ in the
$6P_{3/2}$ and $6P_{1/2}$ excited state manifolds, and where $P_3$ and
$P_4$ are projection operators onto the $F=3$ and $F=4$ ground state
manifolds.
The quantity $\Delta_{HF} \equiv (2\pi)(9.2~\mathrm{GHz})$ is the
hyperfine splitting between the $F=3$ and $F=4$ ground state
manifolds, and $\omega_e$ is the energy of excited state $e$,
where the zero of energy is taken to be halfway between the two
ground state manifolds.

As in section \ref{ssec:raman-three-level-effective-hamiltonian},
we want to derive the Raman
coupling that results when the atom is trapped within an optical
cavity and one of the cavity modes is driven with a pair of beams that
generate standing-wave fields inside the cavity.
The coupling of the atom to the standing-wave fields is described by
a Hamiltonian $H_R$ that has the same form as equation
(\ref{eqn:raman-pair-three-level}), but with $\hat{\Omega}_\pm$ given by
\begin{eqnarray}
  \hat{\Omega}_\pm =
  \gamma\,(I_\pm/I_{sat})^{1/2}\,\psi(\vec{r})\,
  \hat{\epsilon}_\pm^* \cdot \vec{A}.
\end{eqnarray}
Here $I_\pm$ are the maximum intensities of the two fields,
and $\gamma = (2\pi)(5.2\,\mathrm{MHz})$ and
$I_{sat} = 2.19\,\mathrm{mW/cm^2}$ are
the spontaneous decay rate and saturation intensity for the $6P_{3/2}$
excited state manifold.
We can express $I_{sat}$ as
\begin{eqnarray}
  I_{sat} = (4\pi^2/3)(\gamma/\lambda_{D2}^3),
\end{eqnarray}
where $\lambda_{D2} = 852\,\mathrm{nm}$ is the wavelength of the
$6S_{1/2} \rightarrow 6P_{3/2}$ transition.
The quantities $\hat{\epsilon}_\pm$ are the polarizations of the two
fields, which we will take to be linear and mutually orthogonal.
Thus, the vectors $\{\hat{\epsilon}_+, \hat{\epsilon}_-, \hat{k}\}$
form an orthonormal frame, where  $\hat{k}$ is a unit vector that lies
along the cavity axis.
The vector $\vec{A}$ is an atomic lowering operator, and is defined
by
\begin{eqnarray}
  \label{eqn:vector-rasing-operator}
  \vec{A}^\dagger \equiv
  \sum_{J'} \sum_{m'}\sum_m \sum_{q=-1}^1 \beta_{J'}(F',F)\,
  \langle F',m'|1,q;F,m\rangle\,
  |6P_{J'},F',m'\rangle \langle 6S_{1/2},F,m|\,
  \hat{e}_q^*.
\end{eqnarray}
Here $\langle F',m'|1,q;F,m\rangle$ is the Clebsch-Gordan coefficient
that connects ground state $|F,m\rangle$ to excited state
$|F',m'\rangle$ via polarization $\hat{e}_q^*$,
\begin{eqnarray}
  \label{eqn:circular-polarization-vectors}
  \hat{e}_0 = \hat{z},
  \qquad
  \hat{e}_{\pm 1} =
  \mp \frac{1}{\sqrt{2}}(\hat{x} \pm i \hat{y}),
\end{eqnarray}
is a orthonormal basis of polarization vectors, and
\begin{eqnarray}
  \label{eqn:weighting-factor}
  \beta_{J'}(F, F') =
  -(-1)^{F'} \sqrt{2J'+1} \sqrt{2F + 1}
  \left\{
  \begin{array}{ccc}
    1 & 1/2 & J' \\
    7/2 & F' & F
  \end{array}
  \right\},
\end{eqnarray}
is a weighting factor for transitions between the
$6S_{1/2},F$ and $6P_{J'}, F'$ hyperfine manifolds, where
the quantity in brackets is a $6J$-symbol.

Following the procedure we used in section
\ref{ssec:raman-three-level-effective-hamiltonian},
we can adiabatically eliminate the
excited states to obtain an effective Hamiltonian for the ground
states.
Because the derivation closely parallels the derivation given in
section
\ref{ssec:raman-three-level-effective-hamiltonian}, we will omit the
intermediate steps and simply quote the result:
\begin{eqnarray}
  H_E =
  \frac{1}{2}\Delta_{HF}(P_4 - P_3) + \hat{V}_E +
  (\hat{\Omega}_E + \hat{\Omega}_E^\dagger)\,\cos \delta_R t,
\end{eqnarray}
where
\begin{eqnarray}
  \label{eqn:v-eff-sum}
  \hat{V}_E & = &
  \sum_F \sum_e \frac{1}{4\Delta_e}\,
  P_F\,(\hat{\Omega}_+|e\rangle\langle e|\hat{\Omega}_+^\dagger +
  \hat{\Omega}_-|e\rangle\langle e|\hat{\Omega}_-^\dagger)P_F, \\
  \label{eqn:omega-eff-sum}
  \hat{\Omega}_E & = &
  \sum_e \frac{1}{2\Delta_e}\,
  P_3\,\hat{\Omega}_+|e\rangle\langle e|\hat{\Omega}_-^\dagger P_4.
\end{eqnarray}
Here $\Delta_e \equiv \omega_L - \omega_e$ is the overall detuning of
the Raman pair from excited state $e$.
In the limit that $\Delta_e$ is much larger than the excited state
hyperfine splittings, one can show that \cite{boozer05}
\begin{eqnarray}
  \label{eqn:sum-over-excited-states}
  \sum_e \frac{1}{\Delta_e} A_i|e\rangle\langle e|A_j^\dagger
  & = &
  (1/3)[(2/\Delta_{D2} + 1/\Delta_{D1})\,\delta_{ij} +
    2i(1/\Delta_{D2} - 1/\Delta_{D1})\,\epsilon_{ijk}\,J_k],
\end{eqnarray}
where $\Delta_{D1} \equiv \omega_L - \omega_{D1}$ and
$\Delta_{D2} \equiv \omega_L - \omega_{D2}$ are the overall detunings
of the Raman pair from the cesium $D1$ and $D2$ lines.
It is convenient to express these detunings as
\begin{eqnarray}
  \label{eqn:relate-detunings-to-detuning-parameters}
  1/\Delta_{D1} = -(\lambda_L/2\pi)\,C_{D1}, \qquad
  1/\Delta_{D2} = -(\lambda_L/2\pi)\,C_{D2},
\end{eqnarray}
where
\begin{eqnarray}
  \label{eqn:detuning-parameters}
  C_{D1} \equiv (\lambda_L/\lambda_{D1} - 1)^{-1}, \qquad
  C_{D2} \equiv (\lambda_L/\lambda_{D2} - 1)^{-1}
\end{eqnarray}
are dimensionless parameters.
From equations (\ref{eqn:v-eff-sum}), (\ref{eqn:omega-eff-sum}), and
(\ref{eqn:sum-over-excited-states}), we find that
\begin{eqnarray}
  \hat{V}_E =
  V_E\,|\psi(\vec{r})|^2,
  \qquad
  \hat{\Omega}_E =
  \Omega_E\,|\psi(\vec{r})|^2\,\hat{\Sigma},
\end{eqnarray}
where
\begin{eqnarray}
  \label{eqn:v-eff-cesium}
  V_E & \equiv &
  (\gamma^2/12) ((I_+ + I_-)/I_{sat}) (2/\Delta_{D2} + 1/\Delta_{D1}), \\
  \label{eqn:omega-eff-cesium}
  \Omega_E & \equiv &
  (\gamma^2/)(I_+ I_-/I_{sat}^2)^{1/2} (1/\Delta_{D2} - 1/\Delta_{D1}),
\end{eqnarray}
and
\begin{eqnarray}
  \hat{\Sigma} \equiv 2\,P_3\,\hat{k}\cdot\vec{J}\,P_4
\end{eqnarray}
is an atomic lowering operator that couples Zeeman states in $F=4$ to
Zeeman states in $F=3$.
If we collect these results, make the rotating wave approximation,
and perform a unitary transformation to eliminate the time-dependence,
we can express the effective Hamiltonian as
\begin{eqnarray}
  \label{eqn:effective-hamiltonian-cesium}
  H_E =
  -\frac{\delta}{2} (P_4 - P_3)  - V_E\,|\psi(\vec{r})|^2 +
  \frac{1}{2}\Omega_E\,
  |\psi(\vec{r})|^2\,(\hat{\Sigma} + \hat{\Sigma}^\dagger).
\end{eqnarray}
It is instructive to compare
to the effective Hamiltonian for the three-level model
given in equation (\ref{eqn:effective-hamiltonian}) with
the effective Hamiltonian for the full cesium atom
given in equation (\ref{eqn:effective-hamiltonian-cesium}).
The two Hamiltonians have similar forms, only the operator $\sigma_x$
that coupled ground states $a$ and $b$ has been replaced by the
operator $\hat{\Sigma} + \hat{\Sigma}^\dagger$ that couples Zeeman
states within the ground state manifolds $F=3$ and $F=4$.
In addition, we now have
equations (\ref{eqn:v-eff-cesium}) and (\ref{eqn:omega-eff-cesium})
that allow us to calculate the parameters $V_E$ and $\Omega_E$ in
terms of the intensities of the standing-wave fields.

Following the same reasoning that was used in sections
\ref{sec:raman-raman-three-level} and
\ref{sec:fort-raman-three-level}, we can
use the effective Hamiltonian given in equation
(\ref{eqn:effective-hamiltonian-cesium}) to write down the
total Hamiltonian $H$ for the FORT-Raman and Raman-Raman configurations.
In both cases the total Hamiltonian has the form
$H = H_{ext} + H_{int}$, where
\begin{eqnarray}
  H_{ext} & = &
  \frac{p^2}{2m} + U_\rho\,\sin^2 k_F x, \\
  H_{int} & = &
  -\frac{\delta}{2} (P_4 - P_3)  +
  \frac{1}{2}\Omega_\rho\,
  \cos^2 (k_F x + \alpha)\,(\hat{\Sigma} + \hat{\Sigma}^\dagger).
\end{eqnarray}
Here $U_\rho = U_F\,e^{-2\rho^2/w_F^2}$ and
$\Omega_\rho = \Omega_E\,e^{-2\rho^2/w_F^2}$ are the axial trap depth
and the effective Rabi frequency at radial coordinate $\rho$, and the
parameters $U_F$ and $\Omega_E$ are calculated for the FORT-Raman and
Raman-Raman configurations in the following sections
\ref{sec:fort-raman-cesium} and \ref{sec:raman-raman-cesium}.

\subsection{FORT-Raman configuration}
\label{sec:fort-raman-cesium}

As was discussed in section \ref{sec:fort-raman-three-level}, in the
FORT-Raman configuration the FORT forms one leg of the Raman pair, and a
weak Raman beam is added to form the second leg.
The FORT resonantly drives mode $n_F$ of the cavity, and the Raman
beam drives the same mode as the FORT but is detuned from the cavity
resonance by $|\delta_R| \sim \Delta_{HF}$.

We can obtain expressions for the FORT depth $U_F$ and the effective
Rabi frequency $\Omega_E$ by using equations
(\ref{eqn:v-eff-cesium}) and (\ref{eqn:omega-eff-cesium}),
which relate these quantities to the maximum intensities of the FORT
and Raman beams inside the cavity, together with equation
(\ref{eqn:intracavity-intensity}) from Appendix
\ref{sec:cavity-mode-structure}, which relates these maximum
intensities to the optical powers of the FORT and Raman beams at the
input of the cavity
(note that because the Raman beam is detuned from the cavity resonance,
its coupling into the cavity is suppressed).
We find that
\begin{eqnarray}
  \label{eqn:fort-depth}
  U_F & = &
  (\gamma/24\pi)(\gamma/\kappa_F)(2C^F_{D2} + C^F_{D1})(P_F/P_c), \\
  \Omega_E & = &
  (\gamma/12\pi)(\gamma/\kappa_F)(C^F_{D2} - C^F_{D1})
  (1 + (2\Delta_{HF}/\kappa_F)^2)^{-1/2} (P_R P_F/P_c^2)^{1/2}.
\end{eqnarray}
Here $P_F$ and $P_R$ are the powers of the FORT and Raman beams at the
input of the cavity,
$P_c$ is a reference power that is set by the cavity geometry and is
defined in equation (\ref{eqn:power-c}) of Appendix
\ref{sec:cavity-mode-structure},
$\kappa_F$ is the total energy decay rate for the
FORT mode $n_F$, and $C^F_{D2}$ and $C^F_{D1}$, the detuning
parameters at the FORT wavelength $\lambda_F$, are given by
equation (\ref{eqn:detuning-parameters}).
It is interesting to note that for fixed powers in the FORT and Raman
beams, the effective Rabi frequency $\Omega_E$ monotonically increases
as the cavity decay rate $\kappa_F$ is reduced.

In deriving the expression for $U_F$ given in equation
(\ref{eqn:fort-depth}), we assumed that the detuning of the FORT from
the cesium $D1$ and $D2$ lines was the same for the $F=3$ and $F=4$
ground state hyperfine manifolds.
This is a reasonable approximation, because these detunings are much
larger than the hyperfine splitting $\Delta_{HF}$.
However, because the detuning of the $F=3$ manifold is slightly larger
than the detuning of the $F=4$ manifold, the FORT potential is
slightly weaker for $F=3$.
Thus, the FORT squeezes the two manifolds together, causing a small
reduction in the effective hyperfine splitting.
This effect, which is calculated in Appendix
\ref{sec:differential-stark}, gives a slight position-dependence to
the effective Raman detuning, but this can be neglected for many
applications.

\subsection{Raman-Raman configuration}
\label{sec:raman-raman-cesium}

As was discussed in section \ref{sec:raman-raman-three-level}, in the
Raman-Raman configuration the FORT resonantly drives mode $n_F$ of the
cavity, and pair of Raman beams drives mode $n_R$ of the cavity.
We will assume that the two Raman beams have equal powers $P_R$ and 
are tuned symmetrically about the cavity resonance.

The FORT depth $U_F$ is given by equation (\ref{eqn:fort-depth}), and
we can obtain an expression for the Rabi frequency $\Omega_E$ by using
equation (\ref{eqn:omega-eff-cesium}), which relates the Rabi frequency
to the maximum intensities of the Raman beams inside the cavity,
together with equation (\ref{eqn:intracavity-intensity}) from Appendix
\ref{sec:cavity-mode-structure}, which relates these maximum
intensities to the optical powers of the Raman beams at the
input of the cavity (note that because the Raman beams are detuned
from the cavity resonance, their coupling into the cavity is suppressed).
We find that
\begin{eqnarray}
  \Omega_E =
  (\gamma/12\pi)(\gamma/\kappa_R)(C^R_{D2} - C^R_{D1})
  (1 + (\Delta_{HF}/\kappa_R)^2)^{-1} (P_R/P_c).
\end{eqnarray}
Here $P_c$ is a reference power that is set by the cavity geometry and
is defined in equation (\ref{eqn:power-c}) of Appendix
\ref{sec:cavity-mode-structure},
$\kappa_R$ is the total energy decay rate for the
Raman mode $n_R$, and $C^R_{D2}$ and $C^R_{D1}$, the detuning
parameters at the Raman wavelength $\lambda_R$, are given by
equation (\ref{eqn:detuning-parameters}).
Note that for fixed powers in the Raman beams there is an optimal
cavity decay rate $\kappa_R = \Delta_{HF}$ that maximizes the
effective Rabi frequency $\Omega_E$.

\subsection{Zeeman transitions}

The operator $\hat{\Sigma} + \hat{\Sigma}^\dagger$ that appears in
$H_{int}$ couples individual Zeeman transitions between the $F=3$ and
$F=4$ ground state hyperfine manifolds.
In this section, we calculate the matrix elements for these
transitions.

Let us introduce an arbitrary coordinate system
$\{\hat{x},\hat{y},\hat{z}\}$ and define a set of Zeeman states
$\{|3,m\rangle, |4,m\rangle \}$ relative to this coordinate system.
We can express the unit vector $\hat{k}$ that lies along the cavity
axis as
\begin{eqnarray}
  \hat{k} =
  \cos\phi\sin\theta\,\hat{x} +
  \sin\phi\sin\theta\,\hat{y} +
  \cos\theta\,\hat{z},
\end{eqnarray}
where $\theta$ is the angle between the cavity axis $\hat{k}$ and the
quantization axis $\hat{z}$.
Note that
\begin{eqnarray}
  \hat{\Sigma} =
  2 P_3\,\hat{k}\cdot\vec{J}\,P_4 =
  P_3\,(2J_z \,\cos\theta +
  J_+\,e^{-i\phi}\sin\theta + J_-\,e^{i\phi}\sin\theta)\,P_4,
\end{eqnarray}
where $J_\pm = J_x \pm i J_y = \mp \sqrt{2} J_{\pm 1}$ are angular
momentum raising and lowering operators.
Thus, the state $|3,m\rangle$ couples to states $|4,m\rangle$ and
$|4,m\pm 1\rangle$, and the matrix elements corresponding to these
transitions are
\begin{eqnarray}
  \label{eqn:zeeman-zero}
  \langle 4, m|\hat{\Sigma}^\dagger|3, m\rangle & = &
  (1 - m^2/16)^{1/2}\,\cos\theta, \\
  \langle 4, m+1|\hat{\Sigma}^\dagger|3, m\rangle & = &
  \frac{1}{8}(4 + m)^{1/2}(5 + m)^{1/2}\,e^{-i\phi}\,\sin\theta, \\
  \langle 4, m-1|\hat{\Sigma}^\dagger|3, m\rangle & = &
  \frac{1}{8}(4 - m)^{1/2}(5 - m)^{1/2}\,e^{i\phi}\,\sin\theta,
\end{eqnarray}
where we have used that the matrix elements of $\vec{J}$ are given by
\cite{boozer05}
\begin{eqnarray}
  \langle F_2, m_2 |J_q| F_1, m_1 \rangle =
  -(3/2)^{1/2}\,(-1)^{F_2}\,\sqrt{2F_1 + 1}\,
  \left\{
  \begin{array}{ccc}
    1 & 1/2 & 1/2 \\
    7/2 & F_2 & F_1
  \end{array}
  \right\}\,
  \langle F_2, m_2 | 1, q; F_1, m_1\rangle.
\end{eqnarray}
Note that if the quantization axis is aligned along the cavity axis
then $\Delta m = \pm 1$ transitions are forbidden, and if the
quantization axis is transverse to the cavity axis then
$\Delta m = 0$ transitions are forbidden.

\section{Resolved-sideband cooling}
\label{sec:cooling}

\subsection{Cooling schemes}
\label{ssec:cooling-schemes}

We have shown that the Raman coupling can drive transitions that raise
or lower the axial vibrational quantum number $n$.
In this section, we show how one can exploit these $n$-changing
transitions to cool the axial motion to the vibrational ground state.
We will first show how the cooling works using the three-level model,
and then discuss cooling for a physically realistic cesium atom.

One way to cool the atom is to alternate coherent Raman pulses
tuned to an $n$-lowering transition with incoherent repumping pulses.
To see how this works, let us assume that we start out with the atom
in state $|a,n\rangle$.
We can lower the vibrational quantum number by driving the atom with
a coherent Raman pulse tuned to the $n\rightarrow n-1$ transition,
which transfers some of the population from $|a,n\rangle$ to
$|b,n-1\rangle$.
The atom can then be repumped to ground state $a$ by driving the $b-e$
transition with near-resonant light.
The repumping light drives the atom to the excited state, from which
it spontaneously decays to either ground state $b$, where it continues
to interact with the repumping light, or to ground state $a$, where it
is dark to the light.
If the atom is sufficiently cold to begin with, then the repumping
process is unlikely to change the atom's vibrational state, because
the matrix elements for $n$-changing transitions are suppressed
relative to the matrix elements for $n$-preserving transitions by at
least $\eta_e\sqrt{n}$, where
$\eta_e \equiv (2m\omega)^{-1/2}\,\omega_e$.
Thus, the net effect of the Raman and repumping pulses is to move some
of the population from state $|a,n\rangle$ to state $|a,n-1\rangle$.
By iterating the pulse sequence, the atom can be cooled to a state
that has a mean vibrational quantum number $\bar{n}$ that is close to
zero.

The same type of scheme can be used to cool a multi-level alkali atom.
For a cesium atom, the $F=3$ ground state manifold plays the role of
state $a$ and the $F=4$ ground state manifold plays the role of state
$b$: we start with the atom in a random Zeeman state in $F=3$, drive
the atom with a coherent Raman pulse tuned to the
$n \rightarrow n - 1$ transition, and then repump the atom to $F=3$.
It is easiest to understand the effects of these pulses if we choose
the quantization axis to lie along the cavity axis, so that only
$\Delta m = 0$ transitions are allowed and the Raman coupling drives
transitions between pairs of states
$|3,m\rangle \leftrightarrow |4,m\rangle$.
If the ambient magnetic fields are nulled, then these Zeeman
transitions are all degenerate, so the transition frequency of the
$n \rightarrow n - 1$ transition is independent of $m$.
Thus, the coherent Raman pulse is effective at lowering the
vibrational quantum number regardless of which Zeeman state in $F=3$
the atom started in: each Zeeman pair behaves equivalently, except for
a slight $m$-dependence in the Rabi frequency that is given by
equation (\ref{eqn:zeeman-zero}).
The Zeeman state of the atom is then scrambled during the repumping
phase, so at the beginning of the next cooling cycle the atom starts
out in a potentially new Zeeman state.

The amount of time it takes to cool the atom is determined by the
amount of time it takes to repump the atom, which is set by the
spontaneous decay rate of the excited state, and by the amount of time
it takes to perform the coherent Raman pulse, which is set by the Rabi
frequency $\Omega_{n\rightarrow n-1}$.
The cooling rate can be increased by increasing the Rabi frequency,
but as we increase the Rabi frequency we begin to off-resonantly drive
the $n \rightarrow n$ transition, and this sets an upper limit to the
Rabi frequency that can be used.
Off-resonant driving of the $n\rightarrow n$ transition becomes
important when $\Omega_{n\rightarrow n} \sim \omega$, so the upper
limit to the Rabi frequency is given by
$\Omega_{n\rightarrow n-1} \sim \eta\sqrt{n}\,\omega$.

There is also a lower limit to the value of $\bar{n}$ that can be
achieved with this cooling scheme, which is set by two different
factors.
First, when we resonantly drive the $n \rightarrow n-1$ transition
with the coherent Raman pulse,
we can also off-resonantly drive the $n \rightarrow n+1$ transition.
This mechanism gives a lower limit of
$\bar{n} \sim (\Omega_{0\rightarrow 1}/2\omega)^2$.
We can reduce this limit by reducing the Rabi frequency, but since the
Rabi frequency determines the cooling rate, this also slows down the
cooling.
In addition, there are problems with using small Rabi frequencies that
are due to the anharmonicity of the FORT, which will be discussed
later.
Ideally, one would gradually reduce the Rabi frequency as the atom
cools, so as to balance the conflicting demands for a high cooling
rate and a low value of $\bar{n}$.
A second factor that limits $\bar{n}$ is the fact that when the atom
is repumped it will not always remain in the same vibrational state
it started out in, since the Lamb-Dicke suppression of
$n$-changing transitions is not perfect.
This mechanism gives a lower limit of $\bar{n} \sim \eta_e^2$.

The cooling scheme described above can be modified in several ways.
First, rather than alternating Raman pulses with repumping pulses, it
is also possible to continuously drive the atom with both Raman and
repumping light, and this is the method that was used in
\cite{boca04} and \cite{boozer06}.
Second,  the cooling scheme we described relies on $\Delta n = -1$
transitions, but it is also possible to cool the atom using
$\Delta n = -2$ transitions.
Indeed, for the FORT-Raman configuration the $\Delta n = -1$
transitions are forbidden, so the atom can only be cooled using
$\Delta n = -2$ transitions.
Cooling via $\Delta n = -2$ transitions tends to be slower than
cooling via $\Delta n = -1$ transitions, since
the condition $\Omega_{n\rightarrow n} \sim \omega$ gives an upper
limit on the Rabi frequency of
$\Omega_{n\rightarrow n-2} \sim \eta^2 n\,\omega$.
Also, for $\Delta n = -2$ transitions both the state $|a,0\rangle$ and
the state $|a,1\rangle$ decouple from the Raman pulse, so the state to
which the system cools depends on the initial state: if we start in a
state $|a,n\rangle$ with $n$ even then the system cools to
$|a,0\rangle$, and if we start in state $|a,n\rangle$ with $n$ odd
then the system cools to $|a,1\rangle$.

Note that because the FORT is anharmonic, the resonant frequency of
the $\Delta n = -1$ and $\Delta n = -2$ transitions depends on the
value of $n$.
This means that if we keep the Raman detuning $\delta$ set at at a
fixed value throughout the cooling process, then the detuning of the
Raman pulse from the atom will change as the atom cools.
We can estimate the importance of this effect by considering
$\Delta n = -1$ and $\Delta n = -2$ transitions as separate
cases.
First we will consider $\Delta n = -1$ transitions.
Let us assume that we set the Raman detuning to $\delta= -\omega$, so
the detuning of the Raman pulse from the $n\rightarrow n-1$ transition
is
\begin{eqnarray}
  \Delta_{n\rightarrow n-1} =
  \delta - \delta_{n\rightarrow n - 1} \sim
  \eta^2 n\,\omega.
\end{eqnarray}
As we have shown, the maximum Rabi frequency that can be used is
$\Omega_{n\rightarrow n-1} \sim \eta \sqrt{n}\,\omega$, and for
this maximum value the ratio of the detuning to the Rabi frequency is
$\sim \eta \sqrt{n}$, which is small for cold atoms.
Thus, for $\Delta n = -1$ transitions the dependence of the
detuning on $n$ is a small effect; we can simply set the Raman
detuning to $\delta = -\omega$, and as long as the atoms start out
reasonably cold the cooling will always be efficient.

Now consider $\Delta n = -2$ transitions.
We will assume that the Raman detuning is set to $\delta = -2\omega$,
so the detuning of the Raman pulse from the $n\rightarrow n-2$
transition is
\begin{eqnarray}
  \Delta_{n\rightarrow n-2} =
  \delta - \delta_{n\rightarrow n-2} \sim
  2\eta^2 n\,\omega.
\end{eqnarray}
As we have shown, the maximum Rabi frequency that can be used is
$\Omega_{n\rightarrow n-2} \sim \eta^2 n\,\omega$, and for
this maximum value the ratio of the detuning to the Rabi frequency is
$\sim 2$.
Thus, for $\Delta n = -2$ transitions the dependence of the
detuning on $n$ is a significant effect.
To compensate for this problem, one could slowly decrease the Raman
detuning $\delta$ during the cooling process to ensure that the Raman
pulse remains in resonance as the atom cools.

Although we have focused on cooling the axial motion of the atom,
it is possible to implement the axial cooling schemes in such a way
that they cool the atom's radial motion as well.
This is accomplished by configuring the repumping light so that it
provides polarization gradient cooling  \cite{boiron96} in the plane
transverse to the cavity axis.
Specifically, the repumping light is tuned blue of the
$6S_{1/2}, F=4 \rightarrow 6P_{3/2}, F'=4$ transition, and is
delivered to the atom via two pairs of counter-propagating beams.
The two pairs of beams are perpendicular to one another and to the
cavity axis, and therefore provide cooling in both transverse
directions.

\subsection{Measuring the temperature}

One can characterize the effectiveness of the cooling schemes
described in the previous section by using Raman spectroscopy to
measure the temperature of the atom.
In what follows, we will assume that $\Delta n = -1$ transitions are
used to cool the atom, although the same methods can also be applied
to cooling via $\Delta n = -2$ transitions.

To measure a Raman spectrum, we cool the atom, pump it into ground
state $a$, and then drive it with a coherent Raman pulse with detuning
$\delta$.
We then check if the atom was transfered to $b$.
By iterating this sequence one can measure the probability that the
atom is transfered from $a$ to $b$ by the Raman pulse, and by repeating
this measurement for Raman pulses of different detunings one can map
out a Raman spectrum.
For an atom in vibrational state $n$, the Raman spectrum will exhibit
a peak at $\delta = 0$, which corresponds to $n\rightarrow n$
transitions, and peaks at $\delta = \pm \delta_n$, which
correspond to $n \rightarrow n \pm 1$ transitions.
We will refer to the peak at $\delta = 0$ as the carrier,
and the peaks at $\delta = -\delta_n$ and $\delta = \delta_n$ as the
red and blue sidebands.
Because of the FORT anharmonicity $\delta_n$ depends on $n$, but we
will assume that the atoms are cold enough that this effect can be
neglected and simply take $\delta_n \simeq \omega$.

One way to determine the axial temperature of the atom is to measure
the ratio of the red to the blue sideband; this is the same technique
as was used in \cite{monroe95} to determine the temperature of a
trapped ion.
For a thermal distribution, the probability that the atom has axial
vibrational quantum number $n$ is given by
\begin{eqnarray}
  P_n =
  \frac{1}{\bar{n} + 1}
  \left(\frac{\bar{n}}{\bar{n} + 1}\right)^n,
\end{eqnarray}
where $\bar{n} \equiv (e^{\beta\omega} - 1)^{-1}$ is the mean
vibrational quantum number.
If we start with the atom in state $a$ and resonantly drive the blue
sideband with a Raman pulse of duration $t$, the probability that the
atom is transfered to state $b$ is
given by
\begin{eqnarray}
  p_b = \sum_{n=0}^\infty P_n\,\sin^2(\Omega_{n\rightarrow n+1} t/2).
\end{eqnarray}
If we start in state $a$ and resonantly drive the red sideband, the
probability that the atom is transfered to state $b$ is given by
\begin{eqnarray}
  p_r =
  \sum_{n=1}^\infty P_n\,\sin^2(\Omega_{n\rightarrow n-1} t/2) =
  \sum_{n=0}^\infty P_{n+1}\,\sin^2(\Omega_{n+1\rightarrow n} t/2).
\end{eqnarray}
Note that
\begin{eqnarray}
  P_{n+1} =
  \left(\frac{\bar{n}}{\bar{n} + 1}\right)\,P_n,
\end{eqnarray}
so the ratio of the transfer probabilities for the red and blue
sidebands is
\begin{eqnarray}
  p_r/p_b = \bar{n}/(\bar{n} + 1).
\end{eqnarray}

An alternative way to quantify the cooling is to measure the
population in the vibrational ground state.
This can be accomplished by pumping the atom to state $a$ and
applying a Raman pulse whose detuning is adiabatically swept across
the red sideband.
If the atom started in the vibrational ground state $n=0$ then it will
remain in state $a$, and if the atom started in a vibrational state
$n > 0$ then the Raman pulse will adiabatically transfer it to state
$b$.
Thus, the population in the vibrational ground state is given by the
probability that the atom remains in state $a$ after the adiabatic
sweep has been completed.
The advantage of this method is that it does not rely on assuming that
the atoms are thermally distributed.

It is also possible to use Raman spectroscopy to say something about
the radial temperature: since the axial frequency
$\omega(\rho) = \omega_a\,e^{-\rho^2/w_F^2}$ depends on the radial
coordinate $\rho$, the width of the sidebands depends on the radial
temperature.
The probability that the atom has axial frequency $\Delta$ is given by
\begin{eqnarray}
  p(\Delta) =
  \frac{1}{Z}\int_0^\infty e^{-\beta U(\rho)}\,
  \delta(\Delta - \omega(\rho))\,\rho\,d\rho,
\end{eqnarray}
where $U(\rho) = -U_F\,e^{-2\rho^2/w_F^2}$ is the potential for radial
motion, and
\begin{eqnarray}
  Z = \int_0^\infty e^{-\beta U(\rho)}\,\rho\,d\rho.
\end{eqnarray}
If the radial temperature is small compared to the trap depth ($\beta
U_F \gg 1$), then we can make a harmonic approximation and perform the
integral analytically:
\begin{eqnarray}
  p(\Delta) =
  (1/\omega_a)(2\beta U_F)\,
  \theta(1 - \Delta/\omega_a)\,
  \exp(-2\beta U_F(1 - \Delta/\omega_a)).
\end{eqnarray}
Thus, if the blue sideband has width $\delta \omega$, one can put an
upper limit on the radial temperature of
$k_B T < 2U_F(\delta\omega/\omega_a)$.

\subsection{Cooling simulation}

The cooling schemes discussed in section \ref{ssec:cooling-schemes}
can be simulated on a computer.
We will take the Hamiltonian for the system to be
\begin{eqnarray}
  H = H_{int} + H_{ext} + H_L,
\end{eqnarray}
where $H_{int}$ and $H_{ext}$ are given by equations
(\ref{eqn:h-int}) and (\ref{eqn:h-ext}), and where
\begin{eqnarray}
  H_L =
  -\Delta_P\,|e\rangle\langle e| +
  (\Omega_P/2)\,(|b\rangle\langle e| + |e\rangle\langle b|)
\end{eqnarray}
describes the coupling of the atom to repumping light.
Here $\Omega_P$ is the Rabi frequency of the repumping light and
$\Delta_P$ is the detuning of the light from the $b-e$ transition.
As was discussed in section \ref{ssec:cooling-schemes}, in order to
radially cool the atom we use light that is blue-detuned from the
$4-4'$ transition to repump the atom.
To model this in the simulation, we will assume that the excited state
decays to ground state $a$ at rate $\gamma_a = (5/12)\gamma$ and to
ground state $b$ at rate
$\gamma_b = (7/12)\gamma$, where $\gamma = (2\pi)(5.2\,\mathrm{MHz})$
is the spontaneous decay rate for the $6P_{3/2}$ manifold of cesium,
and the prefactors $5/12$ and $7/12$,  the branching ratios for
spontaneous decay on the $6P_{3/2}, F=4 \rightarrow 6S_{1/2}, F=3$ and
$6P_{3/2}, F=4 \rightarrow 6S_{1/2}, F=4$ transitions, are given by
equation (\ref{eqn:weighting-factor}).
Also, we will take $\Delta_P = (2\pi)(10\,\mathrm{MHz})$ to be the
detuning that optimizes the polarization-gradient cooling.

\begin{figure}
  \centering
  \includegraphics[scale=0.5]{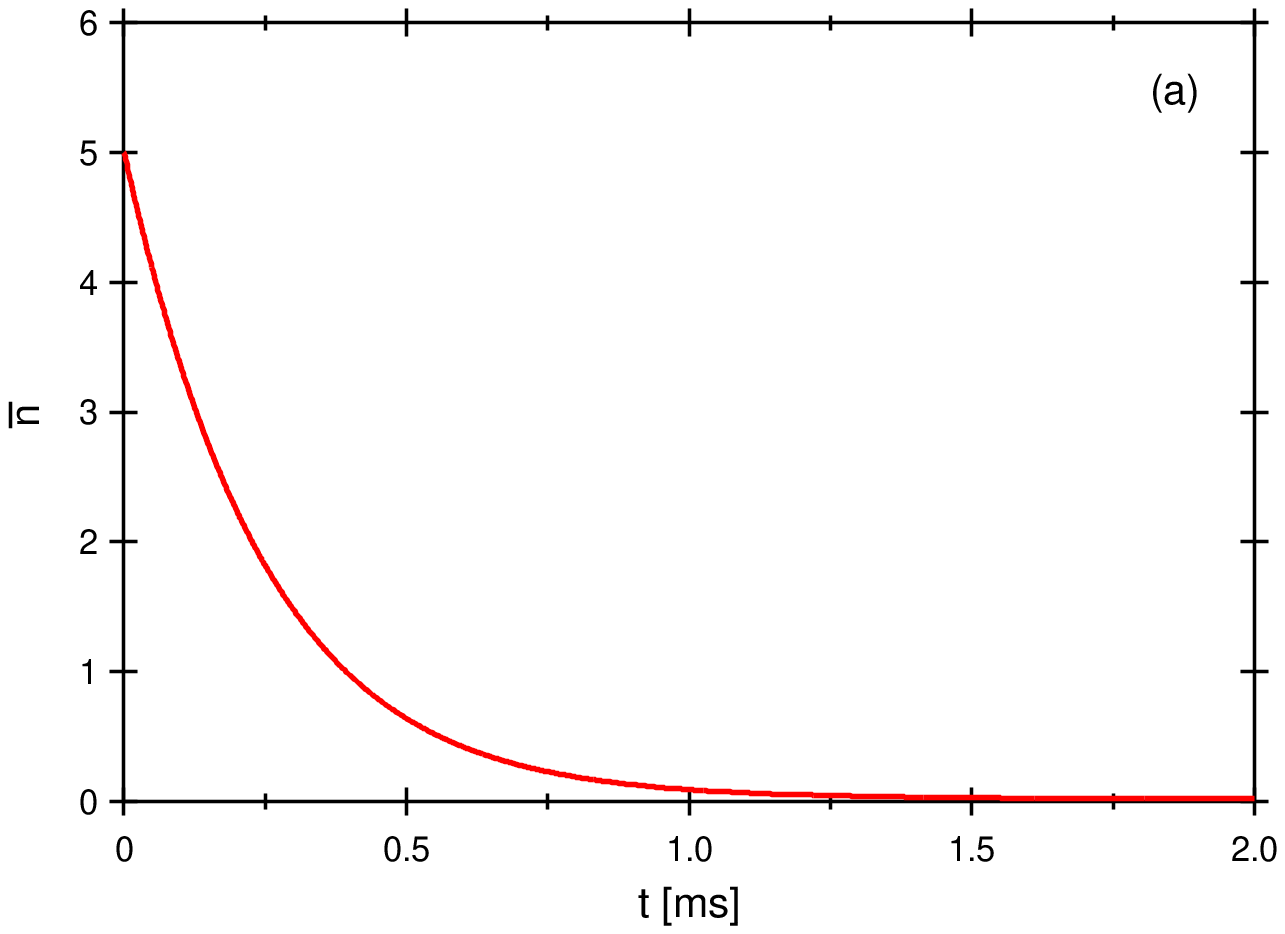}
  \includegraphics[scale=0.5]{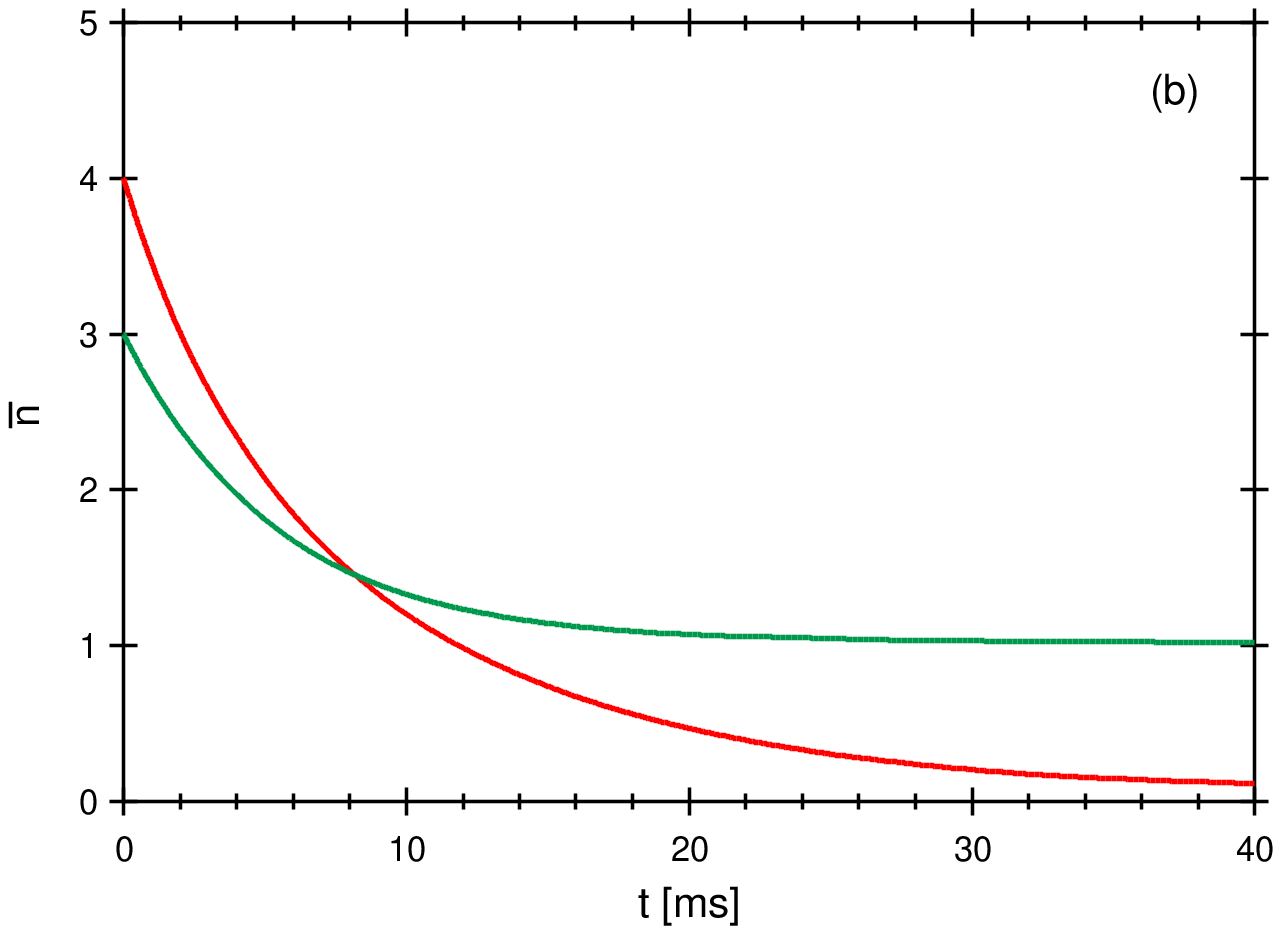}
  \caption{
    Cooling simulation: mean vibrational quantum number
    $\bar{n}$ versus time $t$,
    (a) cooling via $\Delta n = -1$ transitions starting from
    state $|a, 5\rangle$,
    (b) cooling via $\Delta n = -2$ transitions
    starting from states $|a,3\rangle$ and $|a,4\rangle$.
  }
  \label{fig:cooling-1}
\end{figure}

We can write down a master equation for the system, which describes
both the coherent evolution due to $H$ and the incoherent evolution
due to the spontaneous decay from the excited state.
Given an initial state, we can numerically integrate the master
equation to obtain the state of the system at later times.
In Figure \ref{fig:cooling-1}a, we use this method to simulate cooling
in the Raman-Raman configuration: we start the system in state
$|a,5\rangle$ and plot $\bar{n}$ as a function of time.
We assume that the atom is trapped in a FORT well with
$\alpha = \pi/4$, and use $\Delta n = -1$ transitions to perform the
cooling.
For this simulation, the cooling parameters are
$\Omega_\rho = (2\pi)(0.2\,\mathrm{MHz})$,
$\Omega_P = -(2\pi)(3\,\mathrm{MHz})$,
$\delta = -(2\pi)(0.5\,\mathrm{MHz})$.
In Figure \ref{fig:cooling-1}b, we simulate cooling in the FORT-Raman
configuration.
In the FORT-Raman configuration $\alpha = 0$ for all the FORT wells
and the $\Delta n = -1$ transitions are forbidden, so we use
$\Delta n = -2$ transitions to perform the cooling.
As was previously discussed, this means that the asymptotic state to
which the system cools depends on the initial state.
Two curves are shown in the graph: for one, we start the system in
state $|a,3\rangle$; for the other, we start the system in state
$|a,4\rangle$.
For these simulations, the cooling parameters are
$\Omega_\rho = (2\pi)(0.2\,\mathrm{MHz})$,
$\Omega_P = -(2\pi)(3\,\mathrm{MHz})$,
$\delta = -(2\pi)(0.75\,\mathrm{MHz})$.

\begin{figure}
  \centering
  \includegraphics[scale=0.4]{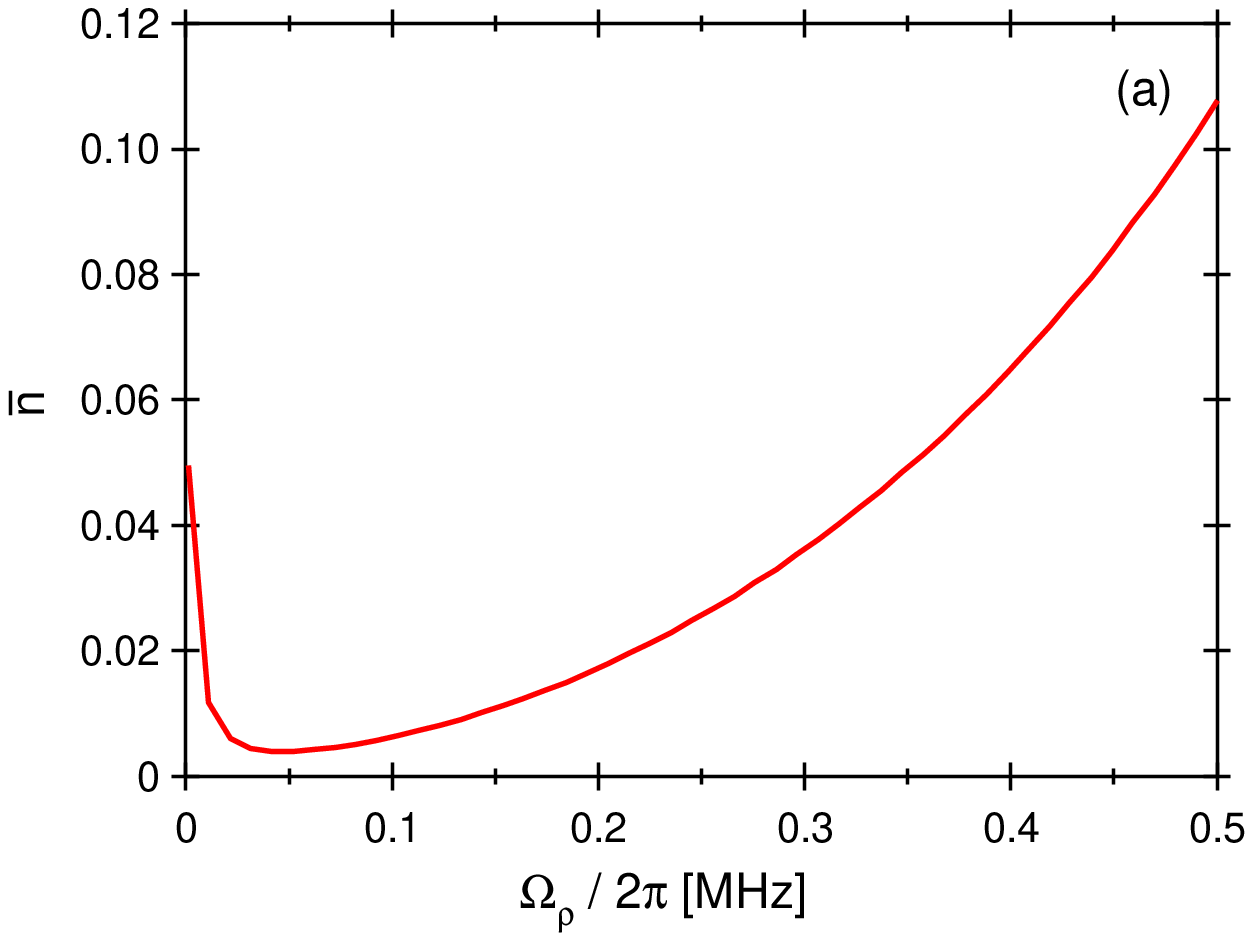}
  \includegraphics[scale=0.4]{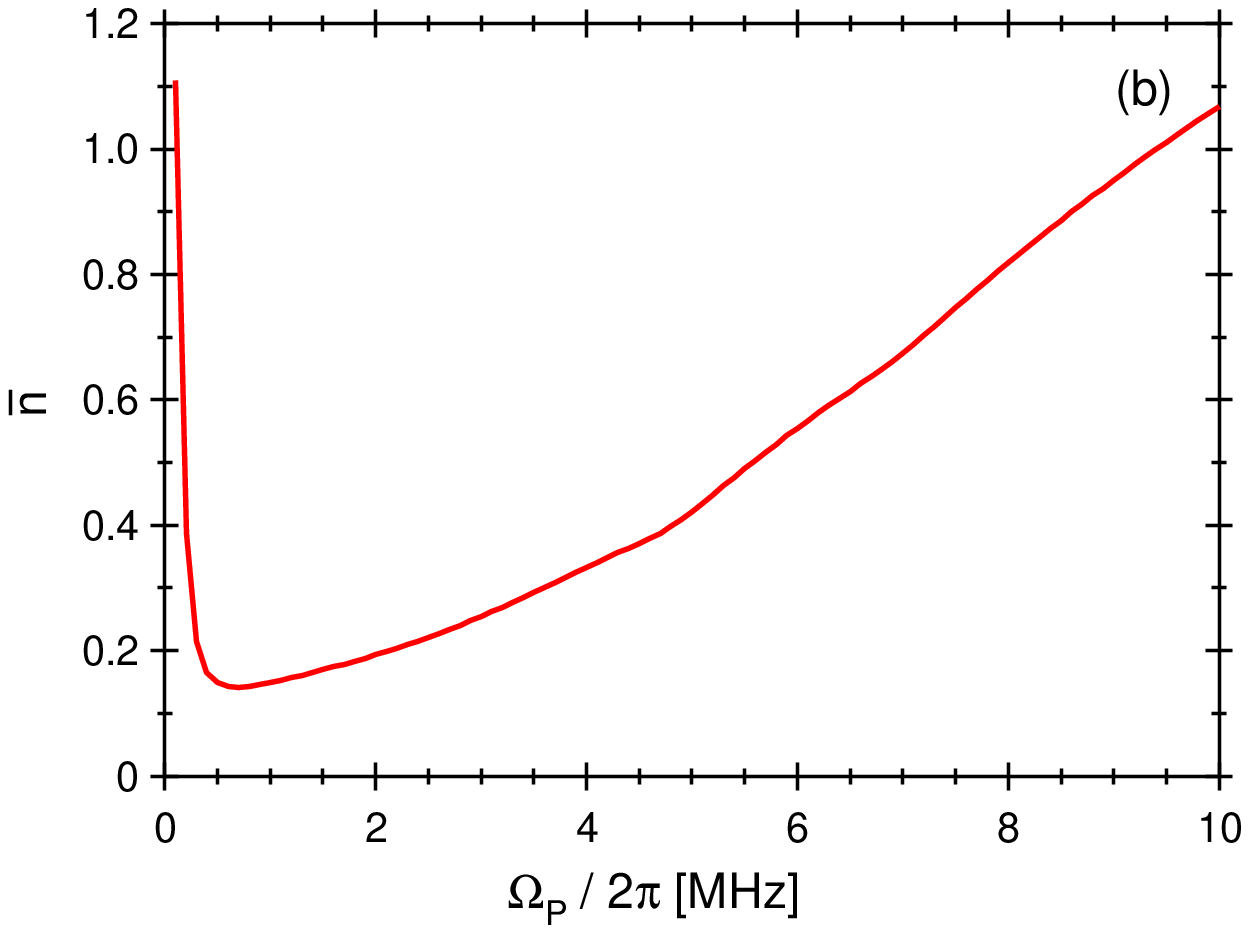}
  \includegraphics[scale=0.4]{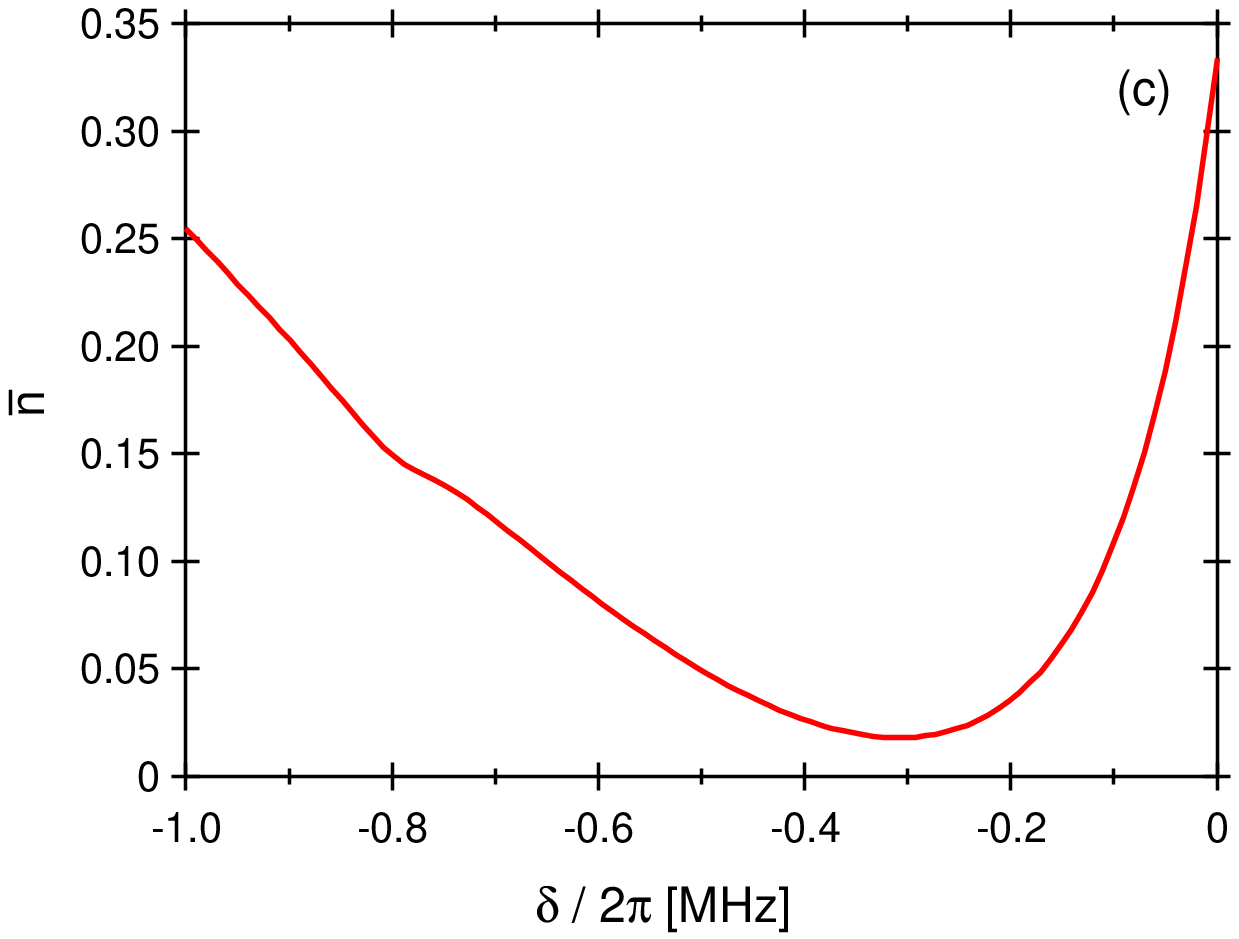}
  \caption{
    Cooling simulation:
    (a) asymptotic $\bar{n}$ versus $\Omega_\rho$,
    (b) asymptotic $\bar{n}$ versus $\Omega_P$,
    (c) asymptotic $\bar{n}$ versus $\delta$.
  }
  \label{fig:cooling-2}
\end{figure}

In addition to simulating the time-evolution of the system, we can
calculate the asymptotic value of $\bar{n}$ by solving the master
equation for the steady-state density matrix.
This can be used to study the dependence of the asymptotic value of
$\bar{n}$ on the various cooling parameters.
In Figure \ref{fig:cooling-2}, we consider cooling in the Raman-Raman
configuration for atoms with $\pi/4$, and plot the asymptotic value of
$\bar{n}$ as a function of $\Omega_\rho$, $\Omega_P$, and $\delta$.
The parameters that are not being varied are set to the same values
used for the cooling simulation shown in Figure \ref{fig:cooling-1}a.
These graphs show that the cooling scheme is quite robust and works
efficiently over a broad range of parameters.

\section{Conclusion}

We have described two schemes for driving Raman transitions in an atom
trapped within a high-finesse optical cavity.
These schemes can be used to control both the internal and motional
degrees of freedom of the atom, and provide powerful tools for
studying cavity QED; as an example, we have shown in detail how the
Raman schemes can be used to cool the atom to the quantum ground state
of the trapping potential.
Although the two schemes are similar in many respects, they do have
some important differences.
The FORT-Raman scheme has the advantage that the Raman coupling is
independent of the FORT well in which the atom is trapped, and is thus
better suited for manipulating the internal state of the atom.
On the other hand, the Raman-Raman scheme has the advantage that the
$n \rightarrow n\pm 1$ transitions are allowed for most FORT wells,
and is thus better suited for cooling.
The ability to coherently control the atom is a key requirement for
many cavity QED protocols, and these Raman schemes should open up new
possibilities for experiments in cavity QED.

\acknowledgements

The author would like to thank A. Boca, R. Miller, and T. E. Northup
for helpful suggestions.
This research is supported by the National Science Foundation,
the Army Research Office, and the Disruptive Technology Office of the
Department of National Intelligence.

\appendix

\section{Cavity mode structure}
\label{sec:cavity-mode-structure}

Here we describe the mode structure of the optical cavity.
The cavity we will be considering consists of two symmetric
mirrors of radius $R$ that are separated by a distance $L$.
It is convenient to define a cylindrical coordinate system that is
centered on the cavity axis: we will denote the distance from the
cavity axis by $\rho$ and the displacement along the cavity axis by
$z$, where the mirrors are located at $z=0$ and $z=L$.
The cavity supports a set of discrete modes with resonant frequencies
at integer multiples of the free spectral range $\nu_{FSR} = 1/2L$,
where for each frequency there are two degenerate modes corresponding
to the two polarization states transverse to the cavity axis.
Consider one of the polarization modes with mode order $n$.
We will let $\omega = 2\pi n \nu_{FSR}$ denote the resonant frequency
of the mode, and let $\lambda = 2\pi/\omega$ and $k = 2\pi/\lambda$
denote the corresponding wavelength and wavenumber.
We can characterize the shape of the mode by a dimensionless function
$\psi(\vec{r})$ that is given by
\begin{eqnarray}
  \label{eqn:mode-shape-general}
  \psi(\vec{r}) = e^{-\rho^2/w^2}\,\sin k z,
\end{eqnarray}
where $w = (L(2R - L)/k^2)^{1/4}$ is the mode radius.
If we drive the cavity with an input beam that has power $P_i$ and
frequency $\omega_i$, then the intensity at a point $\vec{r}$ 
inside the cavity is
\begin{eqnarray}
  I(\vec{r}) =
  (2/\kappa V)(1 + (2\Delta/\kappa)^2)^{-1}\,|\psi(\vec{r})|^2\,P_i,
\end{eqnarray}
where $\Delta = \omega_i - \omega_c$ is the detuning of the input beam
from the cavity resonance, $\kappa$ is the total energy decay rate for
the mode, and $V$, the mode volume, is given by
\begin{eqnarray}
  V = \int |\psi(\vec{r})|^2\,d^3r = (\lambda L/8) (2RL)^{1/2}.
\end{eqnarray}
In order to relate the input power to the maximum intensity inside the
cavity, it is convenient to define a power
\begin{eqnarray}
  \label{eqn:power-c}
  P_c \equiv (2V/\lambda)I_{sat} = (L/4)(2RL)^{1/2}\,I_{sat}.
\end{eqnarray}
Note that $P_c$ is the same for all cavity modes; it depends only on
the cavity geometry, not the mode number.
We can then express the maximum intensity inside the cavity as
\begin{eqnarray}
  \label{eqn:intracavity-intensity}
  I_{max} = (\kappa \lambda)^{-1}\,(1 + (2\Delta/\kappa)^2)^{-1}\,
  (P_i/P_c)\,I_{sat}.
\end{eqnarray}

\section{Distortion of the trapping potential}
\label{sec:potential-distortion}

In the Raman-Raman configuration, the lack of registration between the
FORT and Raman beams causes a well-dependent distortion of the
trapping potential.
Here we calculate this effect.
From equation (\ref{eqn:h-ext-raman-raman}), we see that the total
potential for the Raman-Raman configuration is given by
\begin{eqnarray}
  U(\vec{r}) =
  - U_F\,e^{-2\rho^2/w_F^2}\,\sin^2 k_F z
  - U_R\,e^{-2\rho^2/w_R^2}\,\sin^2 k_R z.
\end{eqnarray}
We will assume that an atom is trapped in well $r$ of the FORT, and
define a coordinate $x = z - z_r$ and a phase
$\alpha = (k_R - k_F)z_r$.
Note that
\begin{eqnarray}
  \sin^2 k_F z = \cos^2 k_F x,
  \qquad
  \sin^2 k_R z = \cos^2(k_R x + \alpha).
\end{eqnarray}
We will assume that the FORT and Raman beams drive nearby modes of the
cavity, so $|k_R - k_F| \ll k_R,k_F$.
In this limit, we can approximate $U(\vec{r})$ by replacing $w_R$ with
$w_F$ and replacing $k_R x$ with $k_F x$:
\begin{eqnarray}
  U(\vec{r}) =
  - U_F\,e^{-2\rho^2/w_F^2}\,\cos^2 k_F x -
  U_R\,e^{-2\rho^2/w_F^2}\,\cos^2 (k_F x + \alpha).
\end{eqnarray}
As in section \ref{sec:trapping}, we will take the atom to be radially
stationary, and take $\rho$ as a constant parameter that enters into
the potential for axial motion.
We can then write the potential as
\begin{eqnarray}
  U(\vec{r}) = U_\rho\,\sin^2 (k_F x + \theta),
\end{eqnarray}
where $U_\rho \equiv U_0\,e^{-2\rho^2/w_F^2}$ is the axial trap depth
at radial coordinate $\rho$.
The quantity $U_0$ is given by
\begin{eqnarray}
  U_0 = (U_F^2 + 2 U_F U_R \cos 2\alpha + U_R^2)^{1/2},
\end{eqnarray}
and $\theta$ is given by
\begin{eqnarray}
  \tan 2\theta = (U_F + U_R\cos 2\alpha)^{-1} U_R\sin 2\alpha.
\end{eqnarray}

\section{Differential Stark shift}
\label{sec:differential-stark}

The FORT potential is slightly weaker for the $F=3$ ground state
manifold than for the $F=4$ ground state manifold, so there is a small
differential Stark shift.
Here we calculate this effect.
The Hamiltonian that describes the differential Stark shift is
\begin{eqnarray}
  H_D =
  \frac{1}{2}\delta_D\,|\psi_F(\vec{r})|^2\,(P_4 - P_3) =
  H_D = \frac{1}{2}\delta_\rho \,(P_4 - P_3)\cos^2 k_F x,
\end{eqnarray}
where $\delta_D$ is the differential Stark shift at an intensity
maximum and $\delta_\rho \equiv \delta_D\,e^{-2\rho^2/w_F^2}$ is the
maximum differential Stark shift at radial position $\rho$.
We can calculate $\delta_D$ as follows.
The FORT depth $U_F$ is given by (\ref{eqn:v-eff-cesium}) with
$I_+ = I_F$ and $I_- = 0$:
\begin{eqnarray}
  U_F =
  (\gamma^2/12)(I_F/I_{sat})(2/\Delta_{D2}^F + 1/\Delta_{D1}^F).
\end{eqnarray}
Thus, the differential Stark shift at an intensity maximum is given by
\begin{eqnarray}
  \delta_D =
  \frac{\gamma^2}{12} \frac{I_F}{I_{sat}}
  \left(\frac{2}{\Delta_{D2}^F} + \frac{1}{\Delta_{D1}^F}\right) -
  \frac{\gamma^2}{12} \frac{I_F}{I_{sat}}
  \left(\frac{2}{\Delta_{D2}^F + \Delta_{HF}} +
  \frac{1}{\Delta_{D1}^F + \Delta_{HF}}\right).
\end{eqnarray}
Expanding to first order in $\Delta_{HF}$, we find that
\begin{eqnarray}
  \delta_D =
  -U_F\,(2(C_{D2}^F)^2 + (C_{D1}^F)^2)(2C_{D2}^F + C_{D1}^F)^{-1}\,
  (\Delta_{HF}/\omega_F).
\end{eqnarray}
where $C_{D1}^F$ and $C_{D2}^F$, the detuning parameters at the FORT
wavelength $\lambda_F$, are given by equation
(\ref{eqn:detuning-parameters}).

\end{document}